\documentclass[12pt, draftclsnofoot, onecolumn]{IEEEtran}

\usepackage{graphicx}
\usepackage{amsmath,bbm,epsfig,amssymb,amsfonts, amstext,verbatim,amsopn,cite,subfigure,multirow,multicol,lipsum}
\usepackage{amsfonts}
\usepackage{epsfig}
\usepackage{epstopdf}
\usepackage{setspace}
\usepackage{stmaryrd}
\usepackage{psfrag}	
\usepackage{multirow}
\usepackage{float}
\usepackage[process=auto]{pstool}
\usepackage{etoolbox}
\usepackage{algorithm}
\usepackage{algorithmic}
\usepackage[export]{adjustbox}
\usepackage{pifont}
\allowdisplaybreaks

\usepackage{clipboard}
\newclipboard{IRS_V2_}
\newcommand{\copyablelabel}[1]{\label{#1}} 

\usepackage{xcolor}
\makeatletter
\let\myorg@bibitem\bibitem
\def\bibitem#1#2\par{%
	\@ifundefined{bibitem@#1}{%
		\myorg@bibitem{#1}#2\par
	}{%
		\begingroup
		\color{\csname bibitem@#1\endcsname}%
		\myorg@bibitem{#1}#2\par
		\endgroup
	}%
}
\newcommand{\highlightref}[1]{\expandafter\newcommand\expandafter*\csname bibitem@#1\endcsname{blue}}
\makeatother

\newtheorem{lem}{Lemma}
\newtheorem{remk}{Remark}

\newtheorem{prop}{Proposition}
\newtheorem{corol}{Corollary}

\EndPreamble
\begin{document}

\title{Intelligent Reflecting Surfaces for Free~Space~Optical~Communication~Systems
\vspace{-0.1cm}}
\author{
	Marzieh Najafi, Bernhard Schmauss, and Robert Schober
\thanks{This paper was presented in part at IEEE Globecom 2019 \cite{Globecom2019}.}
}

\maketitle
\vspace{-1.5cm}
\begin{abstract} 
In this paper, we investigate the use of intelligent reflecting surfaces (IRSs) to relax the line-of-sight requirement of free space optical (FSO) systems. Considering a Gaussian laser beam, we first design a phase-shift distribution across the IRS that enables the reflection of the incident beam in any desired direction, i.e., realizing the generalized law of reflection. Moreover, for the designed phase-shift profile, we show that there exists an equivalent mirror-assisted FSO system that generates a reflected electric field on a mirror that is identical to that on the IRS in the original system. However, the location of the laser source and the properties of the emitted Gaussian laser beam are different  in the original and the equivalent systems. This equivalence allows us to study the mirror-assisted system, employing the image method from  geometric optics, instead of directly analyzing the original IRS-assisted system. Based on this analysis, we model the geometric and misalignment losses (GML) and characterize the impact of the physical parameters of the IRS, such as its size, position, and orientation, on the end-to-end FSO channel. Moreover, we develop a statistical model for the GML which accounts for the random movements of IRS, transmitter (Tx), and receiver (Rx) due to building sway. Furthermore, we analyze the outage probability of an IRS-assisted FSO link based on the derived channel model. Our simulation results validate the accuracy of the developed channel model and offer various insights for system design. For instance, both our simulations and theoretical analysis reveal that even if the variances of the fluctuations of the Tx, IRS, and Rx positions caused by building sway are identical, their impact on the end-to-end channel is not necessarily the same and depends on the relative positioning of these three nodes.
\end{abstract}

\section{Introduction}

\begin{figure*}[t]
	\centering
	\includegraphics[valign=c,width=0.8\linewidth]{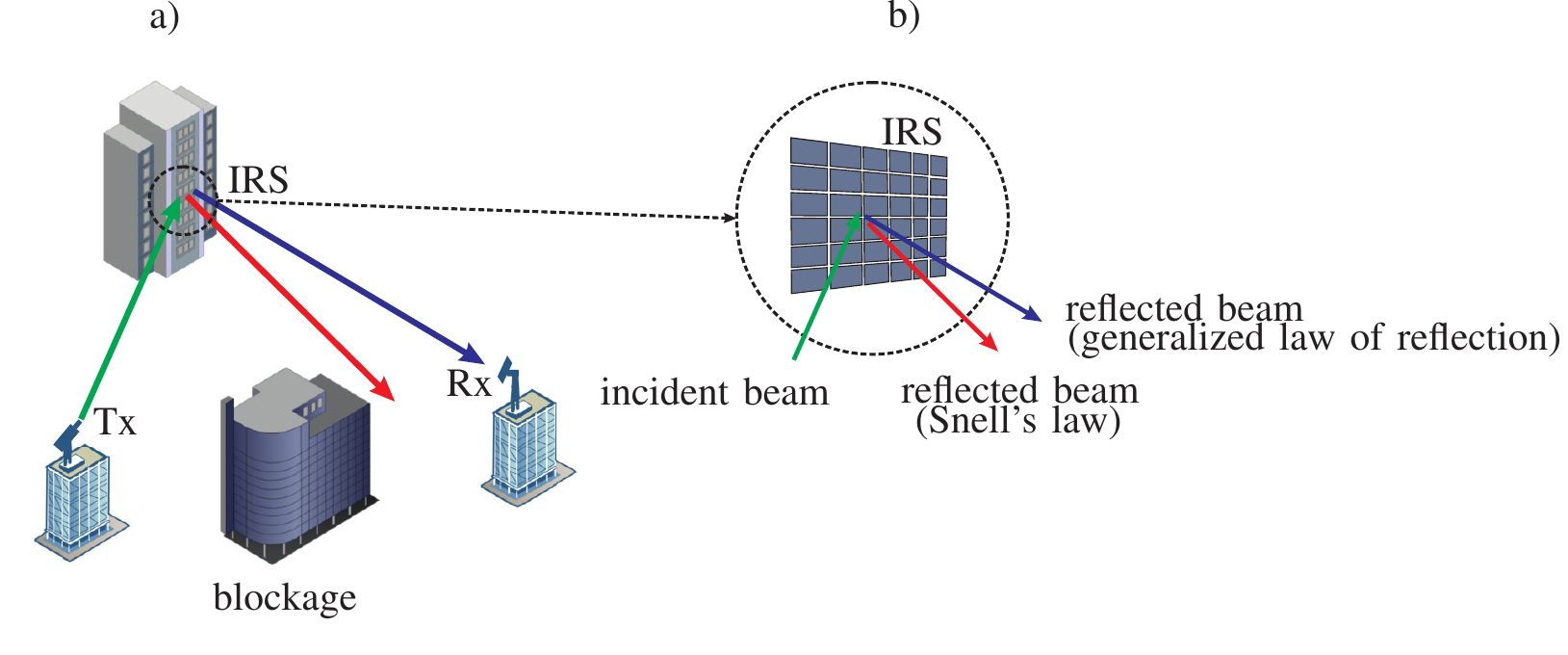}
	\caption{a) IRSs can be deployed to relax the LOS requirement between Tx and Rx in FSO systems. b) IRS can realize anomalous reflection in a desired direction.}
	\label{Fig:IRS}
\end{figure*}

Optical wireless systems, e.g., free space optical (FSO) systems, are promising candidates to meet the high data rate requirements of the next generation of wireless communication networks and beyond \cite{FSO_Survey_Murat,Steve_pointing_error,lee2004part,tsiftsis2009optical}. FSO systems offer the large bandwidth needed for high rate applications, such as  video surveillance and wireless backhauling, while the corresponding transceivers are relatively cheap compared to their radio frequency (RF) counterparts and easy to implement \cite{FSO_Survey_Murat}. However, FSO systems require a line-of-sight (LOS) link between the transmitter (Tx) and receiver (Rx)  and are impaired by
 atmospheric turbulence induced fading, large atmospheric losses in dense fog and heavy snowfall, and geometric and misalignment losses (GML) \cite{FSO_Survey_Murat}. To overcome these impairments, various techniques have been proposed in the literature to improve the reliability of FSO systems, including aperture
averaging \cite{Khalighi2009ApertureAveraging}, diversity techniques \cite{lee2004part,tsiftsis2009optical}, adaptive optics \cite{tyson2002bit}, and RF links as backup for FSO links \cite{FSO_Vahid,MyC_RAN}. Nevertheless, the requirement of an LOS link still remains a severe limitation for the applicability of FSO systems. In the absence of an LOS link, a viable solution  is to deploy a relay node that has an LOS to both Tx and Rx \cite{Li2019RelayFSO,MyTCOM}. However, such a relay node has to be equipped with a partial or complete FSO transceiver chain. For example, a decode-and-forward relay node requires a laser source (LS), lens, photo-detector (PD), tracking system, and signal processing units, which significantly increases the system complexity. To avoid this drawback, in this paper, we propose to employ reconfigurable optical intelligent reflecting surfaces (IRSs) to relax the LOS requirement for FSO systems. Unlike relays, IRSs are energy- and cost-efficient since they are composed of passive elements and can be installed on existing infrastructure, e.g., building walls, cf. Fig.~\ref{Fig:IRS} a).  

%

  IRSs are planar arrays of resonant sub-wavelength elements which enable reflection of the incident beam in a desired direction, i.e., anomalous reflection, by manipulating the wavefront by changing the phase of the beam over the surface of the IRS. This corresponds to the generalized law of reflection \cite{minovich2015functional,yu2011light}, cf. Fig.~1 b). In fact, IRSs belong to the larger category of metasurfaces which are able to change the properties of an impinging electromagnetic wave including its amplitude, phase, dispersion, momentum, and polarization \cite{OpticMetaSurf3,OMS_Tun}. Metasurfaces can be classified into two categories, namely reconfigurable surfaces (see, e.g.,  \cite{OMS_Ref,estakhri2014manipulating,OMS_Tun,metamirror}) and non-reconfigurable surfaces (see, e.g., \cite{yu2011light,olivieri2015plasmonic,GSnellLaw,OpticMetaSurf3}), depending on whether or not the configuration of the surface can be changed post fabrication. 
\Copy{Element}{{ The design, fabrication, and analysis of optical metasurfaces have been an active area of research in the nanotechnology, material science, and physics literature, see review papers \cite{Luo_meta_survey2019,OMS_Tun,Choudhury2018material,minovich2015functional}.  In particular, various techniques have been proposed to realize the reconfigurability of the IRS elements whereby the surface impedance is changed via tunable loads \cite{OMS_Ref,olivieri2015plasmonic,OMS_FSO_Chip,cao2019reconfigurable}.   For instance, in \cite{olivieri2015plasmonic}, the metasurface is based on  a plasmonic  metal-oxide-semiconductor whose capacitance is electrically tunable. In \cite{OMS_Ref}, a conducting oxide metasurface is  employed that consists of a gold backplane and an indium tin oxide layer followed by an aluminum oxide layer on which a gold nanoantenna array is patterned. Thereby, the tunability originates from the field-effect modulation of the complex refractive index of the conducting oxide layers.}}
In this paper, we focus on the application of optical IRSs to avoid  the LOS requirement between the Tx and Rx in FSO systems. In the following, we summarize the main contributions of this paper.


\begin{itemize}
\item Considering a Gaussian laser beam, we first specify the phase-shift distribution across the IRS that enables the reflection of the incident beam in a desired direction. For this phase-shift profile, we show that there exists an equivalent mirror-assisted FSO system that generates a reflected electric field on the mirror that is identical to that on the IRS in the original system. However, the location of the laser source and the properties of the emitted Gaussian laser beam are different  in the original and the equivalent systems. This equivalence allows us to study the mirror-assisted system, exploiting the image method from  geometric optics, instead of directly analyzing the original IRS-assisted system.
\item We characterize the end-to-end FSO channel as a function of the area, position, and orientation of the IRS by deriving the GML of the end-to-end link, i.e., the Tx-to-IRS-to-Rx link, which we refer to as conditional GML.  
\item In an IRS-assisted FSO link not only the Tx and Rx may be subjected to random movements caused by building sway, but also the IRS. Therefore, we develop a statistical channel model which accounts for the impact of building sway on all three nodes. In particular, we develop statistical channel models for both two-dimensional (2D) and three-dimensional (3D) systems. In fact, since deriving the channel model for 2D systems is more straightforward and provides more insights, we first develop a 2D model. Then, based on the insights gained from the 2D channel model, we develop a channel model for practical 3D systems. 
\item The conditional and statistical  GML models presented in this paper are also applicable to  mirror-assisted FSO systems, whereby a simple mirror is deployed to reflect the optical beam towards the Rx \cite{Globecom2019}. In particular, for the proposed phase-shift profile, the IRS reduces to a mirror if the Rx is in the direction of specular reflection. We note that a mirror is able to reflect the optical beam towards a desired Rx only by \textit{mechanically} adjusting its orientation whereas an IRS is able to \textit{electrically} control the direction of the reflected beam\footnote{We note that IRSs and mirrors offer several advantages and disadvantages with respect to (w.r.t.) each other. For example, mirrors are simpler to implement than IRSs; however, the need for mechanical movement imposes a challenge for their deployment on, e.g., building walls. Moreover, the reconfigurability of an IRS allows the simultaneous reflection of the incident wave towards multiple Rxs which cannot be realized with a simple mirror.}.
\item We analyze the system performance in terms of the outage probability. Thereby, we account for both the GML and the atmospheric turbulence induced fading. For the former, we adopt the statistical model developed in this paper, and for the latter, we consider both log-normal (LN) and Gamma-Gamma (GG) atmospheric turbulence induced fading to account for weak and moderate-to-strong turbulences \cite{Steve_pointing_error}, respectively.
\item Simulations are used to validate our derivations and to illustrate the impact of the system parameters such as the variance of the fluctuations induced by building sway and the beamwidth on system performance. For example, our results reveal that for a given end-to-end distance, when the building sway of either the Tx or IRS is more severe than that of the Rx, placing the IRS approximately in the middle between the Tx and Rx minimizes the outage probability. However, when the building sway of the Rx is more severe, moving the IRS closer to the Rx leads to a smaller outage probability. 

\end{itemize}

Although IRS-assisted RF communication systems  have been  extensively studied, see e.g., \cite{najafi2020physicsbased,qWu2019IRSbeamform,di2020analytical}, the literature on  IRS-assisted optical communication systems is rather sparse \cite{Globecom2019,valagiannopoulos2019metasurface,abdelhady2020vlc}. To the best of the authors' knowledge, the first communication-theoretical analysis of an IRS-assisted FSO link, including  corresponding conditional and statistical GML models, was reported in \cite{Globecom2019}, which is the conference version of this paper.  In contrast to \cite{Globecom2019}, which studied mechanically tunable mirrors, in this paper, we consider electronically tunable IRSs, which can realize anomalous reflection in any desired direction without mechanical movement.
	Furthermore, unlike this paper, an outage analysis was not included in \cite{Globecom2019}.  Recently, IRS-assisted visible light communication (VLC) systems were considered in \cite{valagiannopoulos2019metasurface,abdelhady2020vlc}. However, the analysis in our paper is fundamentally different from that in \cite{valagiannopoulos2019metasurface,abdelhady2020vlc} due to the inherent differences between FSO and VLC systems. For example, FSO systems employ narrow laser beams whereas VLC systems use non-directional light-emitting diodes. Moreover, the authors of \cite{OMS_FSO_Chip} proposed to employ optical metasurfaces for inter/intra chip communications. Furthermore, in \cite{cao2019reconfigurable}, an optical IRS was designed and experimentally tested to establish a non-LOS optical wireless link for an indoor application. However, the problem of modeling the IRS-assisted channel  was not considered in \cite{OMS_FSO_Chip} and \cite{cao2019reconfigurable}.

The remainder of this paper is organized as follows. The system and channel models are presented in Section II. The conditional GML model and the corresponding statistical channel model are derived in Sections III and IV, respectively, for a 2D system. These models are then extended to a practical 3D system in Section V. The outage probability of an IRS-assisted FSO link is analyzed in Section VI. Simulation
results are presented in Section VII, and conclusions are drawn in Section VIII.

\textit{Notations:} Boldface lower-case and upper-case letters denote vectors and matrices, respectively.  $\mathbb{R}^+$ and $\mathbb{R}$ represent the sets of positive real and real numbers, respectively. Superscript $(\cdot)^{\mathsf{T}}$ and $\mathbbmss{E}\{\cdot\}$ denote the transpose and  expectation operators, respectively, and $\|\mathbf{a}\|$ is the Euclidean norm of vector $\mathbf{a}$. Moreover, $\ln(x)$,  $\mathrm{erf}(x)$,  $I_0(x)$, $K_{\nu}(x)$, ${}_p F_q\left(a_1,\dots,a_l;b_1,\dots,b_q;z\right)$,  and $\Gamma(x)$ represent the natural logarithm, the Gaussian error function, the zeroth order modified Bessel function of the first kind, the $\nu$-th order modified Bessel function of the second kind, the hypergeometric function, and the Gamma function, respectively. Furthermore, $\mathbf{I}_N$ is the $N\times N$ identity matrix, $\mathrm{diag}\{a_1,\cdots,a_n\}$ is a diagonal matrix with $a_i,\,i=1,\cdots,n,$ as main diagonal entries, and $\mathbf{a}\cdot \mathbf{b}$ denotes the inner product of vectors $\mathbf{a}$ and $\mathbf{b}$. Finally,  
 $\mathbf{R}_{\tau}=\begin{bmatrix}
\cos \tau & -\sin \tau\\ \sin \tau& \cos \tau
\end{bmatrix}$  denotes a counter-clockwise rotation matrix with rotation angle $\tau$ and $\mathbf{T}_{\tau}=\mathrm{diag}\{\cos\tau,\,1\}$ is~a~transformation~matrix.

\section{System and Channel Models}

In this section, we first present the system model. Then, we introduce the Gaussian laser beam and the considered FSO channel model.

\subsection{System Model}

We consider an FSO communication system comprising a Tx equipped with an LS emitting a Gaussian beam, an IRS, and an Rx equipped with a lens and a PD. We assume that the LOS between the Tx and the Rx is blocked, and hence, the communication between them is enabled by an IRS which has an LOS to both the Tx and the Rx, cf. Fig.~\ref{Fig:IRS}. The aperture of the LS is directed towards the IRS; the IRS reflects the optical beam that it receives towards the lens of the Rx; and the lens focuses the beam to the PD to collect the optical energy. As is customary for the analysis of optical systems \cite{VLC,di2020analytical,yu2011light,estakhri2014manipulating,estakhri2016wave}, we first adopt a 2D system model. The impact of the position and orientation of the LS, IRS, and Rx as well as building sway on the GML can be conveniently analyzed for the 2D system model, cf. Sections III and IV, and then generalized to a practical 3D system model, cf. Section V. In the following, we define the position and orientation of the LS, the IRS, and the lens for both the 2D and 3D system models, cf. Fig.~\ref{Fig:Model}.

\subsubsection{2D System Model}

\begin{figure}
	\begin{minipage}[c]{0.43\textwidth}
		\centering
	\includegraphics[valign=c,width=0.8\linewidth]{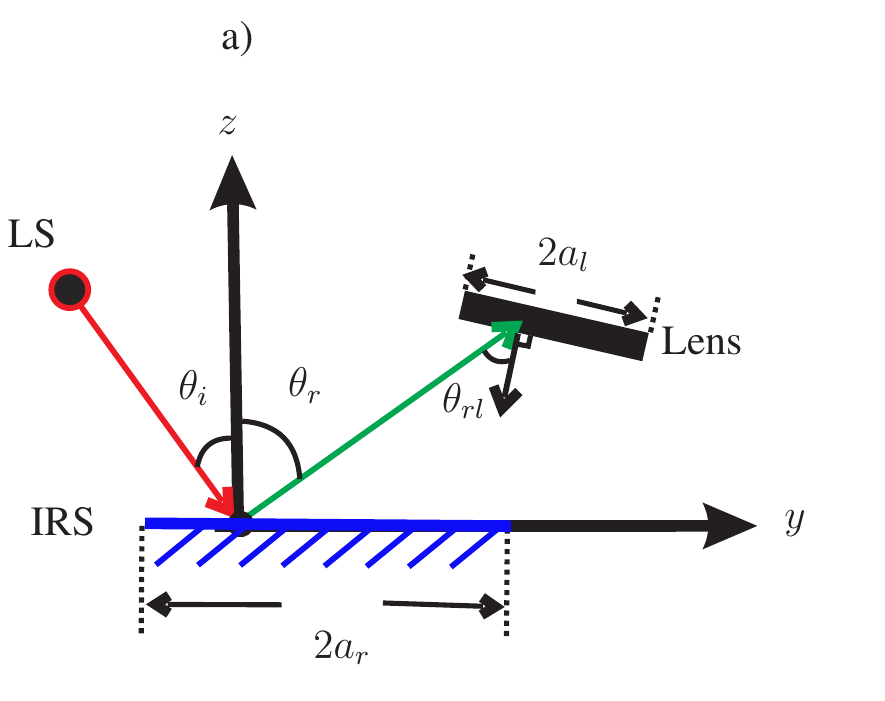}
	\hfill
	\end{minipage} 
	\begin{minipage}[t]{0.06\textwidth}
	\end{minipage}
	\begin{minipage}[c]{0.5\textwidth}
		\centering
		\includegraphics[valign=c,width=0.7\linewidth]{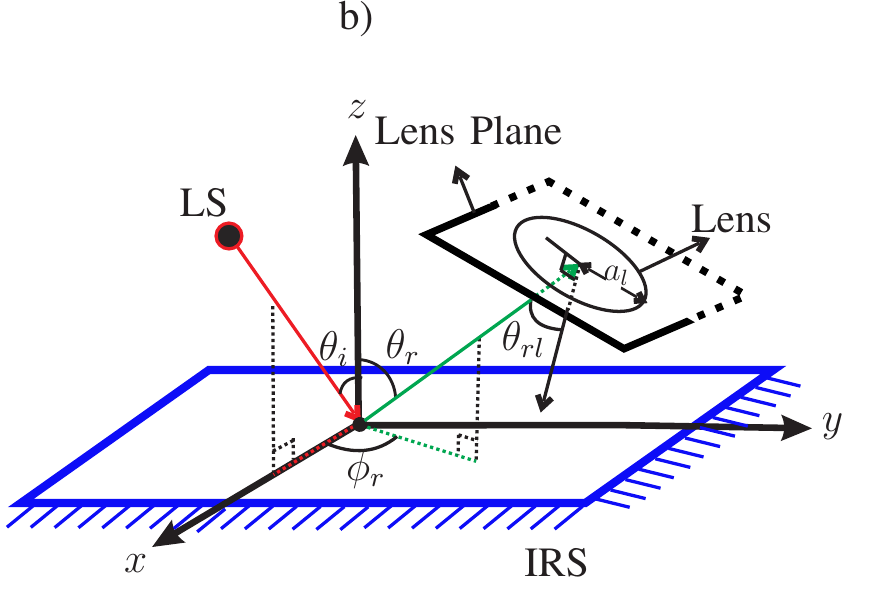}
	\end{minipage}	\vspace{-0.4cm} 
	\caption{Schematic illustration of the considered IRS-assisted FSO system for a) 2D and b) 3D systems.}
	\vspace{-0.4cm}\label{Fig:Model}
\end{figure}

We consider the $yz$-coordinate system and assume that the LS is located at point $(y_{ls},z_{ls})$ and  the centers of the IRS and the Rx lens are located at points $(y_r,0)$ and $(y_l,z_l)$, respectively. Moreover, without loss of generality, we assume that the IRS lies on the $y$ axis and define the origin as the intersection of the beam line\footnote{The beam line is the line that connects the laser source with the center of the beam footprint.} and the IRS line, see Fig.~\ref{Fig:Model}a). The lengths of the IRS line and the lens line
 are denoted by $2a_r$ and $2a_l$, respectively. Moreover, $\theta_i$ and $\theta_r$ denote the angles of incident and reflection at the IRS, respectively, see Fig.~\ref{Fig:Model}a). Note that unlike normal mirrors, for IRSs, $\theta_i$ and $\theta_r$ are not necessarily identical \cite{yu2011light,GSnellLaw}, see also Sections III and V. Finally, let $\theta_{rl}$ denote the angle between the reflected beam and the normal vector of the lens. 

\subsubsection{3D System Model}
We assume that the LS is placed at point $(x_{ls},y_{ls},z_{ls})$, the IRS lies in the $x-y$ plane and its center is at point $(x_r,y_r,0)$, and the lens is a circle with radius $a_l$ centered at point $(x_l,y_l,z_l)$. Similar to the 2D system model, we define the beam footprint center on the IRS as the origin. 
Let $\boldsymbol{\Psi}_i=(\theta_i,\phi_i)$ and $\boldsymbol{\Psi}_r=(\theta_r,\phi_r)$ represent the angles of incident and reflection at the IRS, respectively, where $\theta_m,\,\,m\in\{i,r\}$, denotes the elevation angle (i.e., the angle between the beam line and the $z$ axis) and $\phi_m,\,\,m\in\{i,r\}$, is the azimuth angle (i.e., the angle between the projection of the beam line on the $x-y$ plane and the $x$ axis), see Fig.~\ref{Fig:Model}b). Without loss of generality, we define the $x$ and $y$ axes such that $\phi_i=0$ holds. To characterize the orientation of the lens, we use the angle between the reflected beam and the normal vector of the lens and denote it by $\theta_{rl}$.

\subsection{Gaussian Beam}

For convenience of notation, for presenting the electric field of a Gaussian beam, we define new $\hat{y}\hat{z}$-2D and $\hat{x}\hat{y}\hat{z}$-3D Cartesian coordinate systems, which have the LS located at the origin. A Gaussian beam propagating in $+\hat{z}$ direction has the following electric field~\cite{FourierOptic}
\begin{IEEEeqnarray}{lll}\label{Eq:E}
	E(a,\hat{z})=E_0\Big(\frac{w_0}{w(\hat{z},w_0)}\Big)^{\frac{n-1}{2}}\exp\Big(\frac{-a^2}{w^2(\hat{z},w_0)}\Big)\exp\big(j\phi_{\mathrm{G}}(a,\hat{z},w_0)\big)\quad\text{with phase}\quad	
\end{IEEEeqnarray}
\begin{IEEEeqnarray}{lll}\label{Eq:phase_E}
\phi_{\mathrm{G}}(a,\hat{z},w_0)=-k\hat{z}-k\frac{a^2}{2R(\hat{z},w_0)}+\psi(\hat{z},w_0),
\end{IEEEeqnarray}
where $n$ denotes the dimension of the space, i.e., $n=2$ and $n=3$ for the 2D and 3D spaces, respectively, and $a$ denotes the radial distance from the beam line, which is given by $a=|\hat{y}|$ and $a=\sqrt{\hat{x}^2+\hat{y}^2}$ for 2D and 3D systems, respectively. 
Moreover, $E_0$ denotes the electric field at the origin, $w_0$ denotes the beam waist radius,  $k=\frac{2\pi}{\lambda}$ is the wave number, and $\lambda$ is the optical wavelength. Furthermore, $R(\hat{z},w_0)=\hat{z}\big(1+\left(\frac{z_R}{\hat{z}}\right)^2\big)$ is the curvature radius of the beam's wavefront at distance $\hat{z}$, $z_R=\frac{\pi w^2_0}{\lambda}$, $\psi(\hat{z},w_0)=\frac{n-1}{2}\tan^{-1}\big(\frac{\hat{z}}{z_R}\big)$, and $w(\hat{z},w_0)$ is the beamwidth at distance $\hat{z}$ given by
\begin{IEEEeqnarray}{lll}\label{Eq:Width}
w(\hat{z},w_0)= w_0\sqrt{1+\Big(\frac{\hat{z}}{z_R}\Big)^2}.
\end{IEEEeqnarray}

 The power density of the Gaussian beam in \eqref{Eq:E} at any point on a  line/plane perpendicular to the propagation direction at distance $a$ from the center of the beam footprint is~given~by~\cite{FSO_Survey_Murat,Steve_pointing_error}
\begin{IEEEeqnarray}{lll} \label{Eq:PowerOrthogonal}
I^{\mathrm{orth}}_{\mathrm{G}}(a|\hat{z},w_0) = I^{\max}_{\mathrm{G}}(\hat{z},w_0)\exp\Big(-\frac{2a^2}{w^2(\hat{z},w_0)}\Big),
\end{IEEEeqnarray}
where $I^{\max}_{\mathrm{G}}(\hat{z},w_0)=\frac{\sqrt{2}}{\sqrt{\pi} w(\hat{z},w_0)}$ and $I^{\max}_{\mathrm{G}}(\hat{z},w_0)=\frac{2}{\pi w^2(\hat{z},w_0)}$ for the 2D and 3D spaces, respectively. Note that $ I^{\max}_{\mathrm{G}}(\hat{z},w_0)$ is a normalization constant that ensures that for any given $\hat{z}$, the integral of $I^{\mathrm{orth}}_{\mathrm{G}}(a|\hat{z},w_0)$ over $\hat{y}$ and $(\hat{x},\hat{y})$  for 2D and 3D systems, respectively, is equal to $1$.


\subsection{Channel Model}

We assume an intensity modulation/direct detection (IM/DD) FSO system.  Thereby, the Tx modulates the data onto the intensity of the laser beam. The Rx is equipped with a lens that focuses the optical power striking it on a PD which measures the intensity of the received optical signal, see \cite[Section~III-B]{FSO_Survey_Murat}, \cite[Chapter~2]{ghassemlooy2019optical}, and \cite[Section~II]{TCOM_2020} for more detailed discussions of IM/DD FSO systems. Moreover, we assume that background noise is the dominant noise source at the PD and therefore the noise is independent from the signal \cite{FSO_Survey_Murat}. The received signal at the PD of the Rx, denoted by $y_s$, is given by
\begin{IEEEeqnarray}{lll}\label{Eq:signal}
	y_s=hx_s+n,
\end{IEEEeqnarray}
where $x_s\in \mathbb{R}^+$ is the transmitted optical symbol (intensity),  $n\in \mathbb{R}$ is zero-mean real-valued additive white Gaussian shot noise with variance $\sigma_n^2$  caused by the ambient light at the PD, and $h\in \mathbb{R}^+$ denotes the end-to-end FSO channel gain from the Tx to the Rx. Moreover, we assume an average power constraint $\mathbbmss{E}\{x_s\}\leq{P}$. 

The FSO channel gain, $h$, is affected by several phenomena and can be modeled as follows \cite{Steve_pointing_error} 
\begin{align} \label{Eq:channel}
	h=\eta h_p h_a h_g,
\end{align}
where $\eta$ is the responsivity of the PD and $h_p$, $h_a$, and $h_g$ represent the atmospheric loss, atmospheric turbulence induced fading, and GML, respectively. In particular, the atmospheric loss, $h_p$, is deterministic and represents the power loss over a propagation path due to absorption and scattering of the light by particles in the atmosphere \cite{FSO_Survey_Murat}. 
 The atmospheric turbulence, $h_a$, is a random variable (RV) and induced by inhomogeneities in the temperature and the pressure of the atmosphere. It is typically modeled as LN and GG distributed RV for weak and moderate-to-strong turbulence conditions, respectively \cite{Steve_pointing_error}. The GML, $h_g$, is caused by the divergence of the optical beam along the propagation distance as well as the misalignment between the laser beam line and the center of the Rx lens due to building sway\cite{FSO_Survey_Murat,FSO_Vahid}. We note that the GML is analogous to the free-space path-loss in RF communication systems. Our goal is to mathematically determine the impact of the IRS on the quality of the FSO channel. More specifically, the impact of the IRS on the end-to-end FSO channel is reflected in $h_p$ and $h_g$ as discussed in the following:

\textit{i) Quality of reflection:} In addition to reflection, practical IRSs may also absorb or scatter some fraction of the beam power. Let $\zeta$ denote the reflection efficiency, i.e., the fraction of power reflected by the IRS, which depends on the operating frequency as well as the bias voltage applied to the surface. For FSO systems with $\lambda=1550$~nm,  values in the range $[0.7, 1]$ are typical \cite{OMS_Ref,metamirror}. The absorption at the IRS can be regarded as a part of the atmospheric loss $h_p$. In particular, $h_p$~is~modeled~as~\cite{FSO_Vahid}
		\begin{align} \label{Eq:Pathloss}
			h_p=\zeta10^{-\kappa d_{sr}/10}10^{-\kappa d_{rl}/10} = \zeta10^{-\kappa d_{e2e}/10},
		\end{align}
		where $\kappa$ denotes the weather-dependent attenuation coefficient of the FSO link, $d_{sr}$ and $d_{rl}$ denote the distances from the LS to the IRS and from the IRS to the Rx, respectively, and $d_{e2e}=d_{sr}+d_{rl}$ is the end-to-end distance that the optical beam travels from the Tx to the Rx. As can be observed from \eqref{Eq:Pathloss}, the  atmospheric loss $h_p$  of  the end-to-end IRS-assisted link can be factorized into a product of three terms, namely the  atmospheric loss of the Tx-to-IRS link $10^{-\kappa d_{sr}/10}$, the IRS reflection efficiency $\zeta$, and  the  atmospheric loss of the IRS-to-Rx link $10^{-\kappa d_{rl}/10}$. 


\textit{ii) Relative position, orientation, and size of IRS:} The relative position and orientation of the IRS w.r.t. the laser beam determines the distribution of the \textit{reflected} optical power in space. Moreover, the size of the IRS determines which part of the lens is covered by the reflected beam. These parameters affect the \textit{mean} of the GML $h_g$. 

\textit{iii) Building Sway:} Tx, IRS, and Rx are affected by the random movements of the buildings that they are installed on. This further increases the beam misalignment and affects the \textit{statistics} of the GML $h_g$, i.e., building sway causes $h_g$ to be random.
 
Based on the above discussion, quantifying the impact of the IRS on the end-to-end FSO channel reduces to characterizing the corresponding GML $h_g$. 


\subsection{Geometric Optics and Huygens-Fresnel Principle}

 The GML of an IRS-assisted FSO link is defined as the fraction of the Tx laser beam power that is collected by the optical lens at the Rx after reflection by the IRS \cite{TCOM_2020}, i.e., 
	\begin{align}
	h_g = \int_{\mathbf{r}\in\mathcal{A}^{\rm lens}} I^{\rm lens} (\mathbf{r}) \mathrm{d}\mathbf{r}
	= \frac{2}{\pi w_0^2|E_0|^2} \int_{\mathbf{r}\in\mathcal{A}^{\rm lens}}  |E^{\rm lens} (\mathbf{r})|^2 \mathrm{d}\mathbf{r},\label{Eq:GMLint}
	\end{align}
	where $\mathcal{A}^{\rm lens}$ is the set of points on the Rx lens and $I^{\rm lens} (\mathbf{r})$ is the optical power density at point $\mathbf{r}$  on the Rx lens. $I^{\rm lens} (\mathbf{r})$ is the ratio of the optical  intensity (or irradiance) distribution on the Rx lens, denoted by $S^{\rm lens}(\mathbf{r})$, and  the total power of the Gaussian beam in \eqref{Eq:E}, denoted by $P^{\rm G}$, i.e.,   $I^{\rm lens} (\mathbf{r})=\frac{S^{\rm lens}(\mathbf{r})}{P^{\rm G}}$. Here, $S^{\rm lens}(\mathbf{r})=\frac{1}{2\eta_0} |E^{\rm lens} (\mathbf{r})|^2 $ and $P^{\rm G}=\frac{\pi w_0^2 |E_0|^2}{4\eta_0}$, where $E^{\rm lens} (\mathbf{r})$ represents the electric field on the Rx lens and $\eta_0$ is the free-space wave impedance.

Since the direct link between Tx and  Rx does not exist, the electric field on the Rx lens, $E^{\rm lens} (\mathbf{r})$, is caused by the wave reflected from the IRS. The simplest approach for determining $E^{\rm lens} (\mathbf{r})$ is based on the geometric optics approximation, which models the waves reflected from the surface as rays and is known to be accurate at optical frequencies \cite{FourierOptic}. A more rigorous approach is based on the Huygens-Fresnel principle which treats every point on the IRS as a secondary point source emitting a spherical wave \cite{FourierOptic}. Therefore, the electric field on the Rx lens is the superposition of the waves originating from all secondary sources \cite{FourierOptic}, i.e.,
	\begin{align}
	E^{\rm lens} (\mathbf{r}) = \frac{\varsigma}{j \lambda^{\frac{n-1}{2}}} \int_{\mathbf{p}\in\mathcal{A}^{\rm irs}}  E^{\rm irs} (\mathbf{p}) \frac{\exp(jk|\mathbf{r}-\mathbf{p}|)}{|\mathbf{r}-\mathbf{p}|^{\frac{n-1}{2}}} \exp(j\Delta\phi(\mathbf{p})) \mathrm{d}\mathbf{p}, \copyablelabel{Eq:Huygens}
	\end{align}
	where $\mathcal{A}^{\rm irs}$ is the set of points on the IRS,  $E^{\rm irs} (\mathbf{p})$ is the incident electric field at point $\mathbf{p}$  on the IRS, which can be obtained from \eqref{Eq:E}. Moreover, $\varsigma$ is a constant used to ensure that the amount of power reflected by the IRS is smaller than or equal to the power incident on the IRS, i.e., the IRS passivity condition, see \cite[Remark~1]{najafi2020physicsbased}. Furthermore, $\Delta\phi(\mathbf{p})$ denotes the phase shift applied by the IRS in point $\mathbf{p}$.

Although \eqref{Eq:Huygens} can be used to obtain $E^{\rm lens} (\mathbf{r})$ for any phase-shift profile $\Delta\phi(\mathbf{p})$, the resulting integral cannot always be solved in closed form. In fact, even for the special cases for which \eqref{Eq:Huygens} can be solved, the resulting solution is often complicated, see e.g., \cite[Theorem~1]{ajam2020channel}. Such complicated expressions are not amenable to studying the impact of pointing errors on the IRS-assisted FSO link and deriving the corresponding statistical channel model. Therefore, in this paper, instead of following the direct approach in \eqref{Eq:Huygens}, we take a different route to compute $h_g$, which is based on the following insights: 
		\begin{itemize}
		\item We first note that the electric field reflected from a mirror can be also found from \eqref{Eq:Huygens} with $\Delta\phi(\mathbf{p})=\pi,\,\forall \mathbf{p}$ \cite{FourierOptic}. Although \eqref{Eq:Huygens} suggests that the reflected wave is the superposition of the waves scattered from each point on the mirror, in practice, due to the electrically-large size of the mirror at optical frequencies\footnote{A $1$~mm-by-$1$~mm mirror has an effective electric area (i.e., physical area divided by $\lambda^2$) of $4\times10^5$  for an optical wavelength of $1550$~nm.}, the reflected wave can often be modeled as a narrow beam\footnote{Throughout this paper, we use the term \textit{reflected beam} instead of the term \textit{reflected and scattered wave}.} \cite{FourierOptic}. This is the reason why mirror-assisted systems can be accurately analyzed using tools from geometric optics and image theory \cite{Phys}.
		\item In Sections III and V, we show that for a \textit{specific} choice of the IRS phase-shift profile $\Delta\phi(\mathbf{p})$, the beam reflected from the IRS becomes identical to the beam reflected from a mirror provided that the position of the Tx and the type of emitted beam are properly chosen. This enables us to use geometric optics for the equivalent mirror-assisted system to analyze the considered IRS-assisted system.  
	\end{itemize}

We note that although the proposed analysis methodology is applicable only for the proposed choice of $\Delta\phi(\mathbf{p})$, this choice is sufficiently general to enable anomalous reflection in any desired direction, which is the main objective in this paper. Using this analysis methodology, in Sections~III-V, we develop both a conditional GML model that accounts for the position, orientation, and size of the IRS and a statistical GML model that accounts for the random fluctuations of the IRS's position due to building sway. Nevertheless, in the following subsection, we first study the accuracy of the proposed geometric-optics-based analysis methodology for an example anomalous reflection scenario.
		

\subsection{ Accuracy of Analysis for Practical IRSs}
 
\begin{figure}

  \centering
\includegraphics[width=0.8\linewidth]{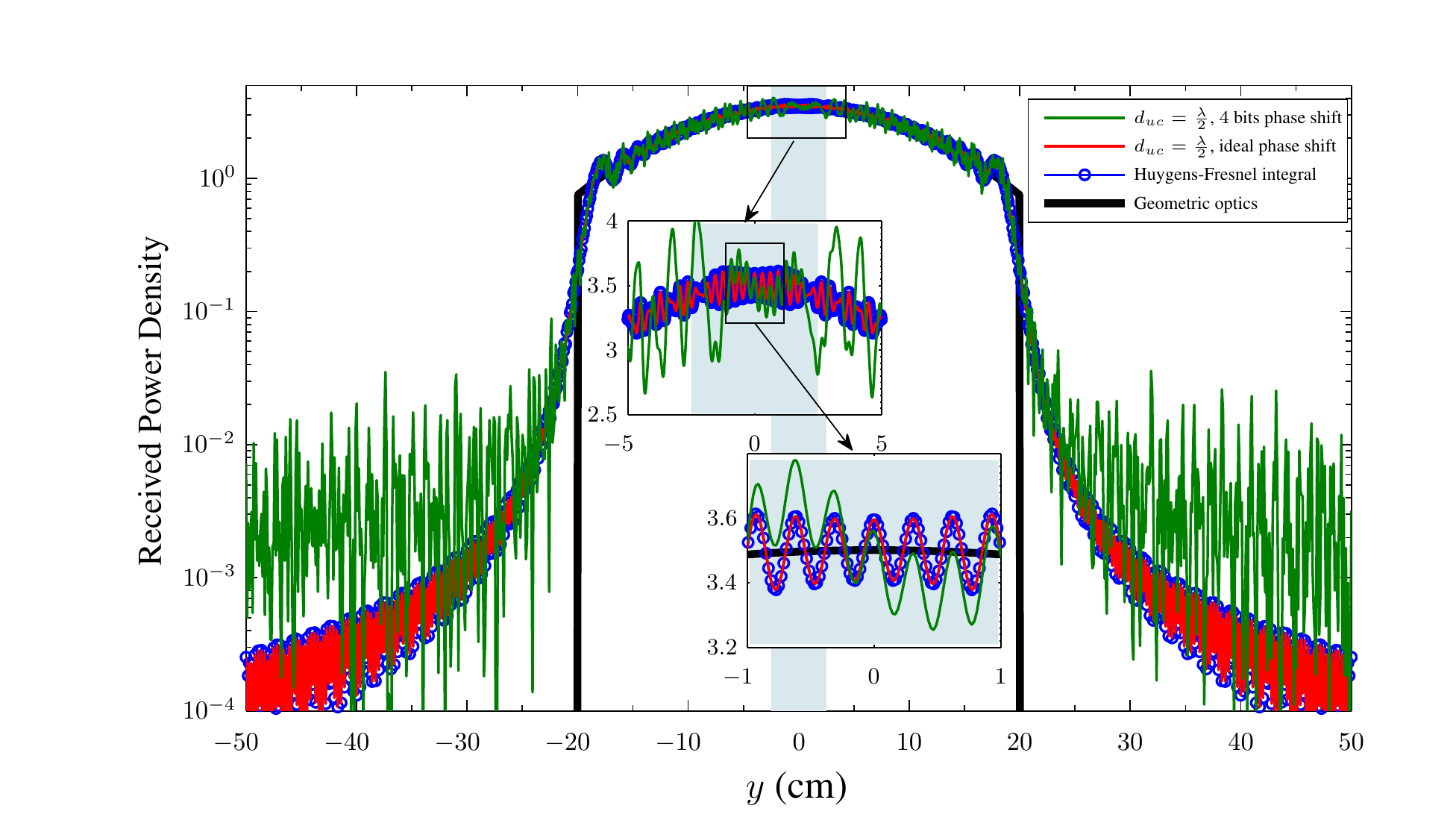}

  \Copy{Fig}{\caption{Received power density over the Rx lens line for the 2D setup in Fig.~2a) with $\theta_i=\frac{\pi}{6}$ rad, $\theta_r=\theta_{rl}=0$, $d_{sr}=d_{rl}=200$~m, $w_0=1$~mm, $\lambda=1550$~nm, $a_r=10$~cm, $a_l=2.5$~cm, $\varsigma=\sqrt{\cos(\theta_i)/\cos(\theta_r)}$, and the proposed phase-shift design in \eqref{Eq:betaxMain}.}}
  \label{Fig:Heugen}

\end{figure}

\Copy{Methods}{ 
The Huygens-Fresnel integral in \eqref{Eq:Huygens} assumes a continuous surface; however, practical optical IRSs consist of discrete sub-wavelength elements \cite{hail2019optical}. Moreover, each element is only able to apply  a finite number of phase-shift values in practice \cite{hail2019optical}, which introduces a phase-shift error. In the following, we compare the power density, $I^{\rm lens} (\mathbf{r})$, obtained when applying \textit{i)} geometric optics, \textit{ii)} the Huygens-Fresnel principle (continuous IRS and ideal continuous phase shift), \textit{iii)} a discrete IRS with continuous phase shifts, and \textit{iv)} a discrete IRS with discrete phase shifts.

Let us consider the 2D system in Fig.~\ref{Fig:Model} a) where the IRS redirects an oblique Gaussian incident wave in the direction perpendicular to the surface via anomalous reflection\footnote{The phase-shift profile $\Delta\phi(\mathbf{p})$  needed to realize anomalous reflection is given \eqref{Eq:betaxMain}.}, where the corresponding system parameters are given in the caption of Fig.~\ref{Fig:Heugen}. In particular, in Fig.~\ref{Fig:Heugen}, we show the power density, $I^{\rm lens} (\mathbf{r})$, vs. $y$  at $z=d_{rl}$, i.e., the Rx lens line. First, we  observe that the received power density computed from the Huygens-Fresnel integral in \eqref{Eq:Huygens} matches well with the power density obtained for a discrete IRS surface with unit-cell element spacing of $d_{uc}=\frac{\lambda}{2}$ and continuous phase-shift values at each IRS element (red curve). Therefore, a discrete IRS surface with  $d_{uc}$ on the order of $\frac{\lambda}{2}$ does not significantly change the received power density compared to that received for a continuous surface. Moreover, taking into account the phase-shift error caused by four-bit quantization of the phase-shift values leads to more variations of power density (green curve).   Furthermore, geometric optics (which will be used to analyze the equivalent mirror-assisted system) is able to accurately predict the power density obtained via the Huygens-Fresnel principle within the range of $|y|< 20$~cm. Outside this range, geometric optics predicts zero  power density due the beam truncation caused by the finite-size IRS, whereas in practice, a small power density exists due to, e.g., diffraction effects. Let us assume an Rx lens with diameter of $5$~cm, which is indicated by the shadowed area in Fig.~\ref{Fig:Heugen}. As can be observed, all four considered cases predict a similar behavior within the lens area. Moreover, since optical lenses are generally very large compared to the wavelength (e.g., $\frac{2a_r}{\lambda}\approx 32,000$ for the considered example), the total  power received by the lens exhibits only a small fluctuation due to the inherent aperture averaging mechanism. For instance, for the example in Fig.~\ref{Fig:Heugen}, the total fraction of the power received by the lens calculated based on geometric optics, the Huygens-Fresnel principle, the discrete IRS with continuous phase-shift values, and the discrete IRS with discrete phase-shift values are $0.339$, $0.340$, $0.340$, and $0.337$, respectively. This reveals an excellent agreement between the results obtained for the four different models.}

Motivated by the accuracy of geometric optics for anomalous reflection, in the following two sections, we use geometric optics and an equivalent mirror-assisted system to derive the conditional and statistical GML models for 2D systems. We generalize these models to practical 3D systems in Section V exploiting the insights gained from the 2D analysis.

\section{Conditional GML - 2D System}
In this section, first, we find the power density distribution of the emitted Gaussian beam across the IRS. Then, we investigate how the IRS changes the direction of the reflected beam, and finally, we determine the conditional GML for  given positions and orientations of the LS, the IRS, and the Rx lens.

\subsection{Power Density Distribution Across IRS} 
 Let $I^{\mathrm{irs}}_{\mathrm{G}}(y|d_{sr},\theta_i,w_0)$ denote the power density distribution across the IRS at point $(y,0)$ which is a function of the position of the LS, specified by $d_{sr}$ and $\theta_i$, and the beam waist radius $w_0$. The following lemma provides the power density distribution across the IRS.

\begin{lem}\label{lem:Lem1} The power density distribution of a Gaussian beam originating from the LS across the IRS is obtained~as
\begin{IEEEeqnarray}{lll} \label{Eq:Iirslem}
I^{\mathrm{irs}}_{\mathrm{G}}(y|d_{sr},\theta_i,w_0) =I^{\mathrm{orth}}_{\mathrm{G}}\left(y|d_{sr},\tilde{w}_0\right),
\end{IEEEeqnarray}
where $\tilde{w}_0$ is the solution of equation $w(d_{sr},\tilde{w}_0)=\frac{w(d_{sr},w_0)}{\cos\theta_i}$ and is given by
\begin{IEEEeqnarray}{lll} \label{Eq:w0new}
\tilde{w}_0=\frac{1}{2}\bigg[\frac{w_0^2}{\cos^2\theta_i}+\frac{\lambda^2 d_{sr}^2}{\pi^2\cos^2\theta_iw_0^2}+\bigg(\Big(\frac{w_0^2}{\cos^2\theta_i}+\frac{\lambda^2 d_{sr}^2}{\pi^2\cos^2\theta_i w_0^2}\Big)^2-4\frac{\lambda^2 d_{sr}^2}{\pi^2}\bigg)^{\frac{1}{2}}\bigg]^{\frac{1}{2}}.
\end{IEEEeqnarray}
\end{lem}
 \begin{IEEEproof}
Please refer to Appendix~\ref{App:I_irs}.
\end{IEEEproof}

Lemma~\ref{lem:Lem1} states that the power density distribution for a non-orthogonal beam impinging on the IRS is equivalent to that of an orthogonal beam originating from a source with the same distance to the IRS, $d_{sr}$, but with a different beam waist radius, $\tilde{w}_0$. Moreover, the width of the orthogonal beam at distance $d_{sr}$ is obtained as $w(d_{sr},\tilde{w}_0)=\frac{w(d_{sr},w_0)}{\cos(\theta_i)}$. Next, we present a useful corollary of Lemma~\ref{lem:Lem1}.


 \begin{corol}\label{corol:corol1} 
 Let us consider a new LS whose location is specified by $\hat{\theta}_i$ and $d_{sr}$ and which emits a Gaussian beam with waist radius $\hat{w}_0$. For any \textit{arbitrary} $\hat{\theta}_i$, the power density distribution across the IRS caused by the new LS becomes identical to that created by the original LS, i.e., 
\begin{IEEEeqnarray}{lll} \label{Eq:ls1ls2I}
I^{\mathrm{irs}}_{\mathrm{G}}(y|d_{sr},\hat{\theta}_i,\hat{w}_0)=I^{\mathrm{irs}}_{\mathrm{G}}(y|d_{sr},\theta_i,w_0),
\end{IEEEeqnarray}
if $\hat{w}_0$ is chosen as the solution to equation $w(d_{sr},\hat{w}_0)=\frac{\cos\hat{\theta}_i}{\cos\theta_i}w(d_{sr},w_0)$ which is given in closed form by \eqref{Eq:w0new} after substituting $\cos\theta_i$ by  $\frac{\cos\theta_i}{\cos\hat{\theta}_i}$.
\end{corol}
\begin{IEEEproof}
According to Lemma~\ref{lem:Lem1}, for the LS,  $I^{\mathrm{irs}}_{\mathrm{G}}(y|d_{sr},\theta_i,w_0)=I^{\mathrm{orth}}_{\mathrm{G}}\left(y|d_{sr},\tilde{w}_0\right)$ holds, where $\tilde{w}_0$ is obtained from $w(d_{sr},\tilde{w}_0)=\frac{w(d_{sr},w_0)}{\cos\theta_i}$. Similarly, for the new LS, $I^{\mathrm{irs}}_{\mathrm{G}}(y|d_{sr},\hat{\theta}_i,\hat{w}_0)=I^{\mathrm{orth}}_{\mathrm{G}}\big(y|d_{sr},\tilde{\hat{w}}_0\big)$ holds, where $\tilde{\hat{w}}_0$ is obtained from $w(d_{sr},\tilde{\hat{w}}_0)=\frac{w(d_{sr},\hat{w}_0)}{\cos\hat{\theta}_i}$. Therefore, $I^{\mathrm{irs}}_{\mathrm{G}}(y|d_{sr},\theta_i,$ $w_0) =I^{\mathrm{irs}}_{\mathrm{G}}(y|d_{sr},\hat{\theta}_i,\hat{w}_0)$ holds when $\tilde{w}_0=\tilde{\hat{w}}_0$ is satisfied. This leads to $\frac{w(d_{sr},\hat{w}_0)}{\cos\hat{\theta}_i}=\frac{w(d_{sr},w_0)}{\cos\theta_i}$, which completes the proof.
\end{IEEEproof}

We use the above corollary in the following subsections to specify a phase-shift profile on the IRS that not only realizes the generalized law of reflection but also allows us to derive the power density distribution of the reflected beam using geometric optics.


\subsection{Phase-shift Profile Across IRS} 
Given the positions of the LS, the IRS, and the lens as well as the properties of the emitted Gaussian beam, we aim to design a phase-shift profile across the IRS, denoted by $\Delta\phi(y|\theta_i,\theta_r,w_0)$, such that it causes the desired angle of reflection $\theta_r$. To this end, we define an equivalent  mirror-assisted system, which is useful for the design of $\Delta\phi(y|\theta_i,\theta_r,w_0)$.

\textit{Proposed Equivalent Mirror-assisted System:} The equivalent system is constructed as follows: \textit{i)} The IRS is replaced by a mirror. \textit{ii)} The LS is replaced by a new LS, denoted by LS$_\mathrm{n}$, which is located in a different position such that it has distance $d_{sr}$ from the mirror and its beam hits the mirror with incident angle $\theta_r$. Therefore, from geometric optics, the angle of reflection at the mirror is $\theta_r$. \textit{iii)} The LS$_\mathrm{n}$ emits a Gaussian beam with waist radius $\hat{w}_0$ given in Corollary~\ref{corol:corol1}. Thereby, according to Corollary~\ref{corol:corol1}, the power density distributions across the mirror and the IRS in both systems are identical. 

\begin{figure}
	\begin{minipage}[t]{0.3\textwidth}
		\centering
		\includegraphics[valign=c,width=0.8\linewidth]{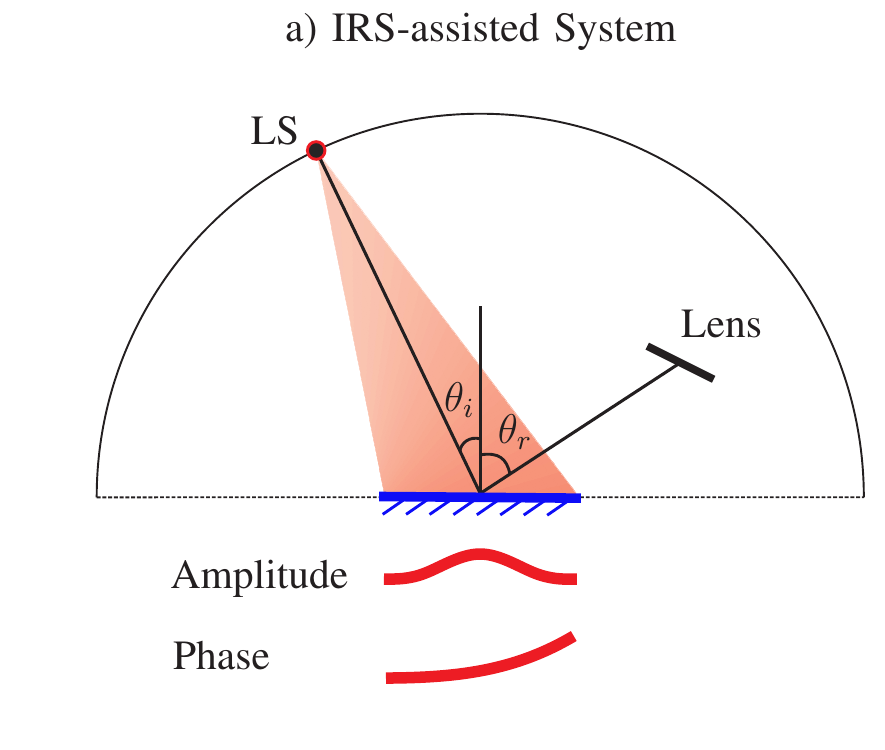}
\vspace{-0.3cm}
	\hfill
	\end{minipage} 
	\begin{minipage}[b]{0.05\textwidth}
	\end{minipage}
	\begin{minipage}[t]{0.3\textwidth}
		\centering
		\includegraphics[valign=c,width=0.8\linewidth]{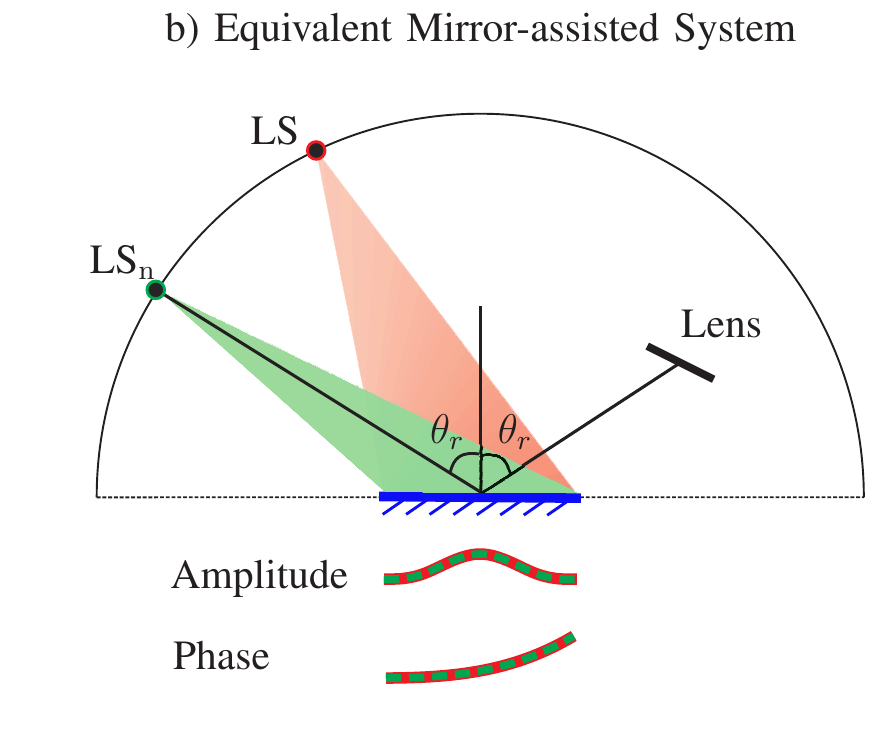}
\vspace{-0.3cm}
	\hfill
	\end{minipage} 
	\begin{minipage}[b]{0.05\textwidth}
	\end{minipage}
		\begin{minipage}[t]{0.3\textwidth}
		\centering
		\includegraphics[valign=c,width=0.8\linewidth]{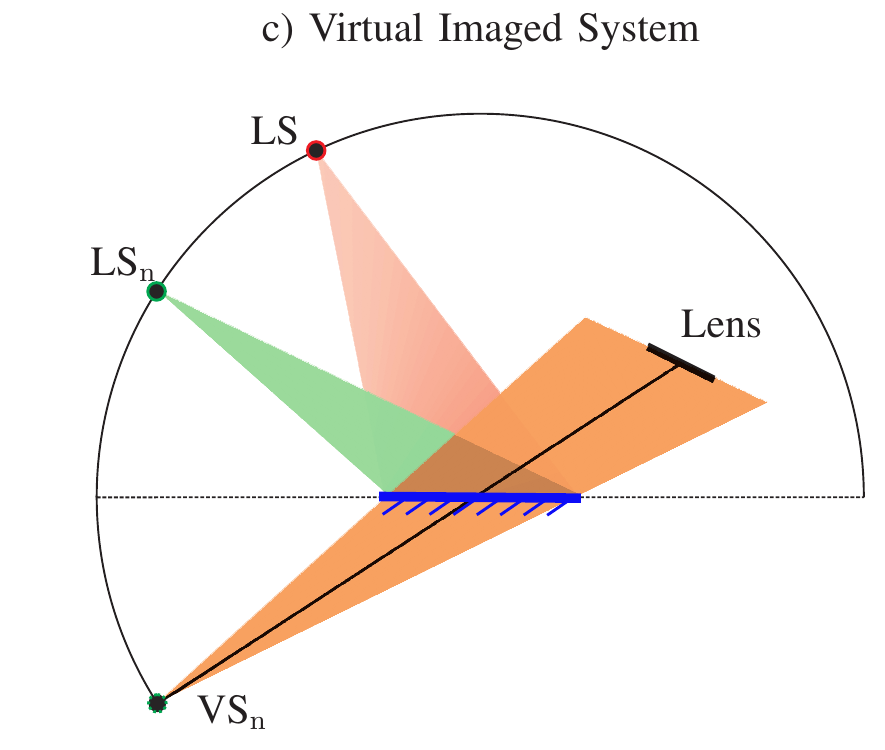}
\vspace{-0.3cm}
	\hfill
	\end{minipage} 
		\caption{Schematic illustration of the proposed equivalent mirror-assisted system. a) Original IRS-assisted system with $\theta_i\neq\theta_r$ and the corresponding amplitude and phase distribution of the reflected beam across the IRS. b) Equivalent mirror-assisted system with $\theta_i=\theta_r$  yielding the same amplitude and phase distribution of the reflected beam across the mirror as those across the IRS in the original system. c) Virtual imaged system.}
	\label{Fig:Equivalent system}\vspace{-0.3cm}
\end{figure}

\textit{Rationale Behind the Proposed Phase-shift Profile:} The location of LS$_\mathrm{n}$ and the properties of its emitted beam (i.e., $\hat{w}_0$) determine the phase distribution of the wave reflected from the mirror. Now, we design the phase-shift profile, $\Delta\phi(y|\theta_i,\theta_r,w_0)$, across the IRS such that the phases of the reflected beams in both systems are identical\footnote{We note that the phase-shift profiles proposed in this paper, cf. \eqref{Eq:betaxMain} and \eqref{Eq:betaxMain_3D}, are not necessarily optimal choices for e.g. maximization of power received at the Rx. However, the reasons for choosing these specific phase-shift profiles are two-folded. On the one hand, the proposed phase shifts realize anomalous reflection in the desired direction which is needed to relax the LOS requirement of FSO systems. On the other hand, they facilitate the use of an equivalent mirror-assisted system for the analysis of the considered IRS-assisted system, which in turn enables the derivation of a closed-form analytical GML model.}. This leads to the same reflected electric fields in both systems. Therefore, the angle of reflection in both systems is also identical and is given by $\theta_r$, cf. Figs.~\ref{Fig:Equivalent system}a), b). 
%


\textit{Proposed Phase-shift Profile:} 
Considering the above discussion, we propose the following phase-shift profile across the IRS to reflect a beam with incident angle $\theta_i$ into direction $\theta_r$
\begin{align}  \copyablelabel{Eq:betaxMain}
\Delta\phi(y|\theta_i,\theta_r,w_0)=\pi+\phi_{\mathrm{G}}(y\cos\theta_r,d_{sr}+y\sin\theta_r,\hat{w}_0)-\phi_{\mathrm{G}}(y\cos\theta_i,d_{sr}+y\sin\theta_i,{w}_0),
\end{align}
where $\phi_{\mathrm{G}}(\cdot,\cdot,\cdot)$ is given in \eqref{Eq:phase_E}. The proposed phase-shift profile reduces to  the well-known constant phase-shift gradient design, i.e., $\Delta\phi(y|\theta_i,\theta_r,w_0)=\pi+ ky\left[\sin\theta_i-\sin\theta_r\right]$,  if $d_{sr}\gg \max\Big\{\frac{\pi w_0^2}{\lambda},$ $\frac{a_r^2|\sin^2\theta_i-\sin^2\theta_r|}{\lambda}\Big\}$ holds\footnote{For $\theta_r\neq \theta_i$ and typical values of $w_0$ and $a_r$ (i.e., $w_0\approx 1$~mm and $a_r\approx 0.01-1$~m), $\pi w_0^2\leq a_{r}^2|\sin^2\theta_i-\sin^2\theta_r|$ holds. Thereby, assuming, e.g., $d_{sr}=1$~km and $\lambda=1550$~nm, the constant phase-shift gradient design is valid  only for $a_{r}\ll 4$~cm. In other words, the constant phase-shift gradient design is not applicable for large IRSs with sizes in the order of a few centimeters.}. The constant phase-shift gradient design has been proposed in the literature to realize the generalized law of reflection for plane waves  \cite{yu2011light,Kochkina2013,estakhri2016wave}. However, in this paper, we consider the Gaussian beam in \eqref{Eq:E} which, unlike plane waves, creates a non-uniform power density distribution as well as a non-linear phase distribution across the IRS, see Fig~\ref{Fig:Equivalent system}a).  To realize the phase-shift profile in \eqref{Eq:betaxMain}, the phase shift introduced by each point on the surface can be related to the corresponding refractive index. For instance, for the IRS design proposed in \cite{OMS_Ref}, a desired phase shift is realized by applying an electrical voltage to the surface that locally changes the refractive index. The phase-shift profile in \eqref{Eq:betaxMain} not only realizes the generalized law of reflection but also enables us to analyze the reflected beam using the equivalent mirror-assisted system.

\begin{remk}\label{Remk:1}
For the special case where $\theta_r=\theta_i$, $\Delta\phi(y|\theta_i,\theta_r,w_0)=\pi$ holds. This corresponds to the $\pi$-phase change that normal mirrors introduce to the reflected beam \cite{Phys}. In other words, for $\theta_r=\theta_i$, the IRS reduces to a simple mirror.
\end{remk}

\subsection{Spatial Distribution of the Reflected Power Density}
As discussed  above, the reflected beams in the IRS- and mirror-assisted systems are identical. Therefore, to find the spatial power density distribution of the reflected beam, we use the mirror-assisted system in Fig~\ref{Fig:Equivalent system}b). It is well known from geometric optics that the system in Fig.~\ref{Fig:Equivalent system}b) with LS$_\mathrm{n}$ and an infinite-size mirror can be transformed to an equivalent system with a virtual LS, denoted by VS$_\mathrm{n}$, without the mirror \cite{Phys}, see Fig.~\ref{Fig:Equivalent system}c). Thereby, VS$_\mathrm{n}$ is the image of LS$_\mathrm{n}$ w.r.t. the mirror. For a mirror of finite size, the lines connecting VS$_\mathrm{n}$ and the boundaries of the mirror create a truncation region, denoted by $\mathcal{R}$, outside of which the reflected optical power is negligible.  The location of VS$_\mathrm{n}$ is obtained as $\mathbf{p}_{vs_\mathrm{n}}=(y_{{vs}_{\mathrm{n}}},z_{{vs}_{\mathrm{n}}})^{\mathsf{T}}=(y_{{ls}_{\mathrm{n}}},-z_{{ls}_{\mathrm{n}}})^{\mathsf{T}}=(-d_{sr}\sin\theta_r,-d_{sr}\cos\theta_r)^{\mathsf{T}}$,
where $(y_{{ls}_{\mathrm{n}}},z_{{ls}_{\mathrm{n}}})$ is the location of LS$_{\mathrm{n}}$. Using these notations, the truncation region can be formally characterized as $\mathcal{R}=\{(y,z)^\mathsf{T}|s_1(y-y_{{vs}_\mathrm{n}})+z_{{vs}_\mathrm{n}}\leq z\leq s_2(y-y_{{vs}_\mathrm{n}})+z_{{vs}_\mathrm{n}}\}$, where $s_1=\frac{-z_{{vs}_{\mathrm{n}}}}{y_r+a_r-y_{{vs}_{\mathrm{n}}}}$ and 
$s_2=\frac{-z_{{vs}_{\mathrm{n}}}}{y_r-a_r-y_{{vs}_{\mathrm{n}}}}$. Moreover, let $\mathbf{p}_{c}=(y_c,z_c)^{\mathsf{T}}$ denote the center of the beam footprint on the lens, where $y_c=\frac{\tan(\theta_r-\theta_{rl})y_l+z_l}{\tan(\theta_r-\theta_{rl})+\cot\theta_r}$ and $z_c=\cot\theta_r y_c$.
Based on the above discussion, the power density of the \textit{beam reflected} by the IRS is formally given in the following lemma.

\begin{lem}\label{Lem:Truncated Gauss}
Let $\hat{\mathbf{p}}=(\hat{y},\hat{z})^\mathsf{T}\in\mathcal{R}$ be a point on a line that is perpendicular to the reflected beam line and passes through $\mathbf{p}_{c}$, cf. Fig.~\ref{Fig:Truncation}.
Assuming the phase-shift profile proposed in \eqref{Eq:betaxMain}, the power density of the beam reflected from the IRS at point $\hat{\mathbf{p}}$ is~given~by
\begin{IEEEeqnarray}{lll} \label{Eq:Truncated Gauss}
I^{\mathrm{orth}}_{\mathrm{rfl}}(\hat{r}|d_{e2e},\hat{w}_0)=
\frac{\sqrt{2}}{\sqrt{\pi} w(d_{e2e},\hat{w}_0)}\exp\Big(-\frac{2\hat{r}^2}{w^2(d_{e2e},\hat{w}_0)}\Big),
\end{IEEEeqnarray}
where  $\hat{r}=\|\hat{\mathbf{p}}-\mathbf{p}_{c}\|$ and $d_{e2e}=\|\mathbf{p}_{vs_\mathrm{n}}-\mathbf{p}_{c}\|$. 
\end{lem}
\begin{IEEEproof}
The reflected beam in the IRS-assisted  system is identical to that in the mirror-assisted system (given in Corollary~\ref{corol:corol1}) which leads to \eqref{Eq:Truncated Gauss}. 
\end{IEEEproof}

Lemma~\ref{Lem:Truncated Gauss} states that the beam reflected by the IRS is a \textit{truncated} Gaussian beam, whose width depends on the angles of incident and reflection and which originates from VS$_\mathrm{n}$ and is confined to area $\mathcal{R}$.  Moreover, the size of $\mathcal{R}$ depends on the size of the IRS as well as on its relative orientation w.r.t. the laser beam.  Note that  $I^{\mathrm{orth}}_{\mathrm{rfl}}(\hat{r}|d_{e2e},\hat{w}_0)$ attains its maximum, i.e., $\frac{\sqrt{2}}{\sqrt{\pi} w(d_{e2e},\hat{w}_0)}$, at $\hat{r}=0$, i.e., at the center of its footprint $\mathbf{p}_{c}$, cf. \eqref{Eq:Truncated Gauss}. Therefore, for an efficient design, we should choose $\theta_r$ such that $\mathbf{p}_{c}$ lies in the center of the lens $(y_l,z_l)$. This leads to an optimal $\theta_r^*$ given by 
\begin{IEEEeqnarray}{lll}\label{Eq:OptimalAngle}
\tan\theta_r^*=\frac{y_l}{z_l}.
\end{IEEEeqnarray}

\subsection{Conditional GML}

We assume that the Rx lens focuses all optical power of the part of the beam that strikes it on the PD. Therefore, in order to compute the fraction of power collected by the PD (i.e., the conditional GML), we have to calculate the fraction of power that flows into the lens. The following proposition provides a closed-form expression for the conditional GML $h_g$. Let $r_c$ denote the distance between the center of the lens and the beam line.

\begin{prop}\label{Prop:TotalFrac}
Under the mild condition $a_l,r_c\ll d_{e2e}$, the conditional GML is given by
\begin{IEEEeqnarray}{lll} \label{Eq:Pt}
h_g=\frac{1}{2}
\begin{cases}\displaystyle
\mathrm{erf}\bigg(\frac{\sqrt{2}\cos\theta_{rl}\rho_1}{w(d_{e2e},\hat{w}_0)}\bigg)+\mathrm{erf}\bigg(\frac{\sqrt{2}\cos\theta_{rl}\rho_2}{w(d_{e2e},\hat{w}_0)}\bigg),\hspace{-2mm}&\mathrm{if}\,\,\rho_{12}=2a_l\\
\displaystyle
\bigg|\mathrm{erf}\bigg(\frac{\sqrt{2}\cos\theta_{rl}\rho_1}{w(d_{e2e},\hat{w}_0)}\bigg)-\mathrm{erf}\bigg(\frac{\sqrt{2}\cos\theta_{rl}\rho_2}{w(d_{e2e},\hat{w}_0)}\bigg)\bigg|,\hspace{-2mm}&\mathrm{otherwise,}
\end{cases}
\end{IEEEeqnarray}
where the term $\cos\theta_{rl}\in[0,1]$ accounts for the non-orthogonality of the lens w.r.t. the beam line\footnote{If there is only one IRS, in order to collect maximum power, it is optimal to adjust the lens such that the reflected beam is orthogonal to the lens. However, if there are more than one IRS, the reflected beams from all IRSs cannot be perpendicular to the lens at the Rx at the same time. Therefore, the Rx needs to employ either a large PD (which would lead to a large field of view at the Rx and in turn a higher background noise level \cite{Safari_AoA}) or more than one PD to collect the powers from the directions of the beams arriving from different IRSs. Therefore, to keep our model general, we assume a non-orthogonal reflected beam which includes the orthogonal beam as a special case.}. Moreover, $\rho_1=\|\mathbf{p}_c-\hat{\mathbf{p}}_1\|$, $\rho_2=\|\mathbf{p}_c-\hat{\mathbf{p}}_2\|$, $\rho_{12}=\|\mathbf{p}_c-\mathbf{p}_1\|+\|\mathbf{p}_c-\mathbf{p}_2\|$, where  $\mathbf{p}_1=(y_l+a_l\cos\tilde{\theta}, z_l-a_l\sin\tilde{\theta})^{\mathsf{T}}$ and $\mathbf{p}_2=(y_l-a_l\cos\tilde{\theta}, z_l+a_l\sin\tilde{\theta})^{\mathsf{T}}$ are the corner points of the lens and  $\hat{\mathbf{p}}_1$ and $\hat{\mathbf{p}}_2$ are given by
\begin{IEEEeqnarray}{ccc} \label{Eq:r_1}
\hat{\mathbf{p}}_1
=
\begin{cases}
\tilde{\mathbf{p}}_2,  \quad &  \mathbf{p}_1^y<\tilde{\mathbf{p}}_2^y\\
\mathbf{p}_1,  &  \tilde{\mathbf{p}}_2^y\leq\mathbf{p}_1^y\leq \tilde{\mathbf{p}}_1^y\\
\widetilde{\mathbf{p}}_1,  &  \mathbf{p}_1^y>\tilde{\mathbf{p}}_1^y
\end{cases},\qquad\quad\,\,
\hat{\mathbf{p}}_2
=
\begin{cases}
\tilde{\mathbf{p}}_2, \quad  &  \mathbf{p}_2^y<\tilde{\mathbf{p}}_2^y\\
\mathbf{p}_2,   &  \tilde{\mathbf{p}}_2^y\leq\mathbf{p}_2^y\leq \tilde{\mathbf{p}}_1^y\\
\tilde{\mathbf{p}}_1,  &  \mathbf{p}_2^y>\tilde{\mathbf{p}}_1^y
\end{cases}\quad\text{with}\nonumber\\
\tilde{\mathbf{p}}_1
=
\begin{bmatrix}
s_1 & -1\\
\tan\tilde{\theta} & 1
\end{bmatrix}^{-1}
\begin{bmatrix}
s_1y_{{vs}_\mathrm{n}}-z_{{vs}_\mathrm{n}}\\
\tan\tilde{\theta}y_l+z_l
\end{bmatrix},\,\,\
\tilde{\mathbf{p}}_2
=
\begin{bmatrix}
s_2 & -1\\
\tan\tilde{\theta} & 1
\end{bmatrix}^{-1}
\begin{bmatrix}
s_2y_{{vs}_\mathrm{n}}-z_{{vs}_\mathrm{n}}\\
\tan\tilde{\theta}y_l+z_l
\end{bmatrix},
\end{IEEEeqnarray}
where $\mathbf{a}^y$ denotes the $y$ component of point $\mathbf{a}$ and $\tilde{\theta}=\theta_r-\theta_{rl}$ is the angle between the lens and the $x$ axis.
\end{prop}
\begin{IEEEproof}
Please refer to Appendix~\ref{App:Prop_TotalFrac}.
\end{IEEEproof}

\begin{figure}
\begin{minipage}[b]{0.47\textwidth}
		\centering
		\includegraphics[valign=c,width=0.8\linewidth]{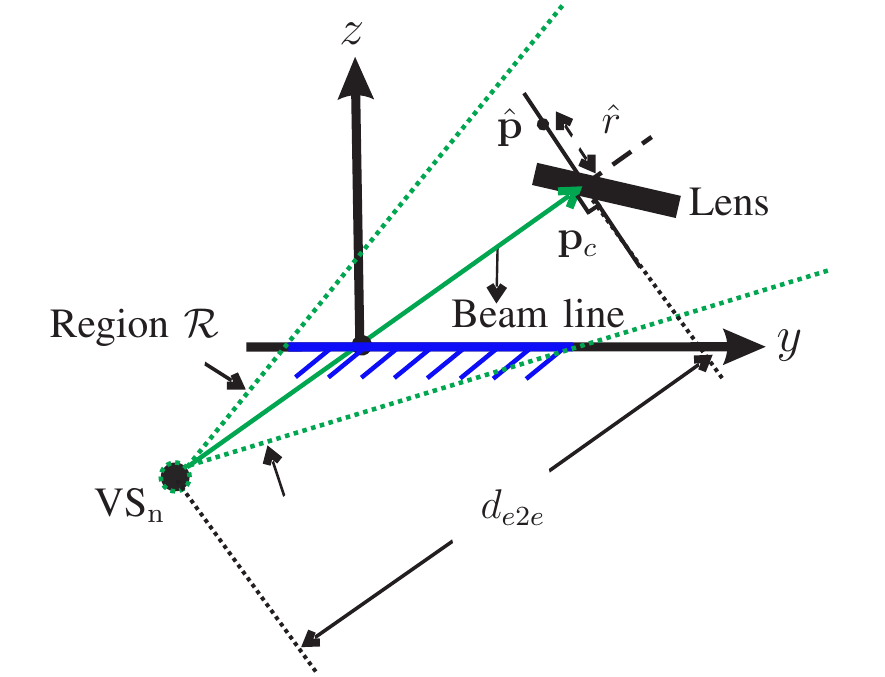}
	\caption{Schematic representation of Lemma~\ref{Lem:Truncated Gauss}.}
	\label{Fig:Truncation}
	\end{minipage} 
		\begin{minipage}[b]{0.06\textwidth}
	\end{minipage}
	\begin{minipage}[b]{0.47\textwidth}
		\centering
		\includegraphics[valign=c,width=0.9\linewidth]{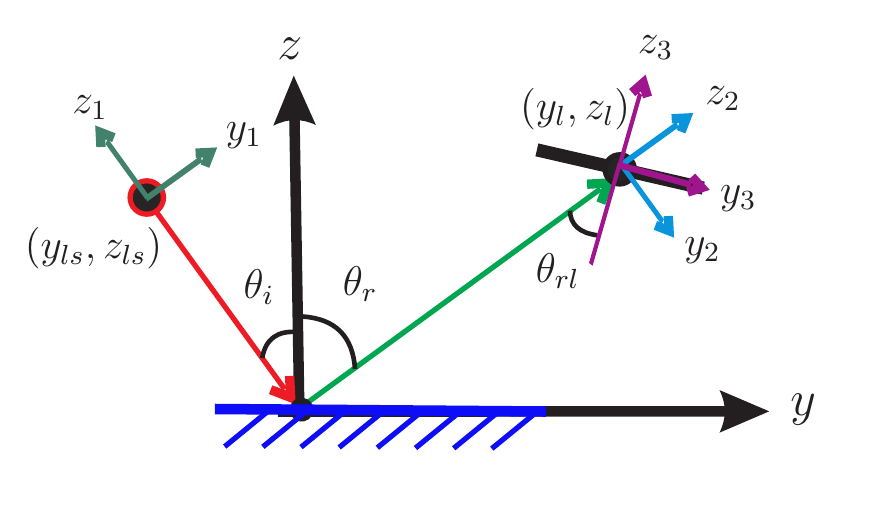}
	\caption{2D coordinate systems for characterization of the LS, IRS, and lens fluctuations as well as the misalignment across the lens line.}
	\label{Fig:Coordinate_Sys}
	\hfill
	\end{minipage} 	
\end{figure}

Note that the conditions under which \eqref{Eq:Pt} in Proposition~\ref{Prop:TotalFrac} holds are met in practice since 1) the physical size of the lens is much smaller than the transmission distance, i.e., $2a_l\ll d_{e2e}$ holds, and 2)  $r_c$ corresponds to the beam misalignment and for a properly designed system, the misalignment is much smaller than the end-to-end transmission distance, i.e.,  $r_c\ll d_{e2e}$ holds. The impact of the size of the IRS is reflected in the values of $\rho_{1}$ and $\rho_{2}$. In fact, if the IRS is sufficiently large such that the lens is located in region $\mathcal{R}$ defined before Lemma~\ref{Lem:Truncated Gauss}, we obtain $\rho_1=\|\mathbf{p}_0-\mathbf{p}_1\|$, $\rho_2=\|\mathbf{p}_0-\mathbf{p}_2\|$, i.e.,  beam truncation does not occur, see Fig.~\ref{Fig:Truncation}. The required IRS size to avoid beam truncation depends on the width of the laser beam at the IRS and the sways of the buildings on which the Tx and the IRS are mounted. Given the typical values for these parameters (e.g., on the order of cm or tens of cm  \cite{Steve_pointing_error}), IRS sizes of a few tens of cm seem sufficient to avoid beam truncation. For larger IRSs, some parts of the IRS are not illuminated with significant optical power, and therefore, do not contribute to the received signal. Considering the increased cost and complexity of larger IRSs, the IRS size should be properly chosen.

\begin{corol}\label{Corol:Perp}
For the special case where $\hat{\mathbf{p}}_i=\mathbf{p}_i, i=1,2$, i.e., the IRS is sufficiently large, and the reflected beam strikes the center of the lens and its direction is perpendicular to the lens, the conditional GML is~given~by
\begin{IEEEeqnarray}{lll} \label{Eq:PtCorol}
h_g=\mathrm{erf}\Big(\frac{\sqrt{2}a_l}{w(d_{e2e},\hat{w}_0)}\Big).
\end{IEEEeqnarray}
\end{corol}
\begin{IEEEproof}
Eq. \eqref{Eq:PtCorol} is obtained by substituting $\theta_{rl}=0$ and $\rho_1=\rho_2=\|\mathbf{p}_0-\mathbf{p}_1\|=\|\mathbf{p}_0-\mathbf{p}_2\|=a_l$ into \eqref{Eq:Pt}. This completes the proof.  
\end{IEEEproof}

For a given lens length $a_l$, the maximum fraction of power collected by the PD is given by \eqref{Eq:PtCorol}. To attain this maximum, three conditions have to hold, namely the IRS is sufficiently large, the misalignment is zero,  i.e., $\theta_r=\theta_r^*$, cf. \eqref{Eq:OptimalAngle}, and the lens is orthogonal to the beam line, i.e., $\theta_{rl}=0$.

\section{Statistical Model - 2D System}
In this section, we study the effect that building sway has on the quality of the considered FSO channel. To this end, we first introduce the considered misalignment model and then derive the probability density function (PDF) of the corresponding GML.

\subsection{Misalignment Model}
We assume that the positions of LS, IRS, and lens fluctuate because of building sway in both the $y$ and $z$ directions.
 In the following, we show that for the LS, IRS, and lens only the fluctuations in a certain direction have a considerable impact on the FSO channel quality, respectively. This observation substantially simplifies the derivation of the proposed statistical GML model. To this end, we define suitable coordinate systems to formally characterize  the fluctuations of LS, IRS, and lens,~cf.~Fig.~\ref{Fig:Coordinate_Sys}.

\textbf{LS:} The fluctuations of the position of the LS can be projected in two directions, one parallel to the beam line, $z_1$, and one orthogonal to it, $y_1$, cf.~Fig.~\ref{Fig:Coordinate_Sys}. Let $\epsilon_s^z$ and $\epsilon_s^y$ denote the fluctuations of the LS position for the former and latter cases, respectively. Hereby, since the fluctuations of the LS in the beam direction are much smaller than the distance between the LS and the IRS, the impact of $\epsilon_s^z$ on $h_g$ can be safely neglected.

\textbf{IRS:} The fluctuations of the position of the IRS can also be projected in two directions, one parallel to the IRS line, $y$, and one orthogonal to it, $z$. Let $\epsilon_r^{y}$ and $\epsilon_r^{z}$ denote the fluctuations along the $y$ and $z$ axes, respectively. Assuming that the beam line is aligned to pass through the IRS (not necessarily its center) and that the size of the IRS is large, the impact of $\epsilon_r^{y}$ on $h_g$ is negligible. 

\textbf{Lens:} Similar to the LS, let $\epsilon_l^z$ and $\epsilon_l^y$ denote the fluctuations of the position of the lens in the direction of the reflected beam, $z_2$, and perpendicular to it, $y_2$, respectively. Since the distance between the IRS and the lens is much larger than the fluctuations in the direction of the reflected beam, we can safely neglect the impact of $\epsilon_l^z$ on $h_g$.

The following lemma provides the misalignment between the center of the beam footprint and the center of the lens, denoted by $u$, in terms of $(\epsilon_s^y,\epsilon_r^z,\epsilon_l^y)$. By convention, we specify $u$ in the direction of axis $y_3$ in Fig.~\ref{Fig:Coordinate_Sys}.
\begin{lem}\label{Lem:Misalignment}
Misalignment $u$  as a function of $(\epsilon_s^y,\epsilon_r^z,\epsilon_l^y)$ is obtained as
\begin{IEEEeqnarray}{lll} \label{Eq:Misalignment}
	u=\frac{1}{\cos\theta_{rl}}\Big(\frac{\cos\theta_r}{\cos\theta_i}\epsilon_s^y - \frac{\sin(\theta_i+\theta_r)}{\cos\theta_i}\epsilon_r^z -\epsilon_l^y\Big).
\end{IEEEeqnarray}
\end{lem}
\begin{IEEEproof}
The proof follows from deriving the contributions of $\epsilon_s^y$, $\epsilon_r^z$, $\epsilon_l^y$ along the $y_3$ axis. For this, we exploit the fact that a misalignment $\hat{a}$ in the line perpendicular to the beam causes misalignment $\tilde{a}=\frac{\hat{a}}{\sin\theta}$ on a line having angle $\theta\in[0,\frac{\pi}{2}]$ with the beam line. Using this result, lens movement $\epsilon_l^y$ along the $y_3$ axis causes misalignment $\frac{-\epsilon_l^y}{\sin(\frac{\pi}{2}-\theta_{rl})}=\frac{-\epsilon_l^y}{\cos\theta_{rl}}$. For the LS,  movement $\epsilon_s^y$ causes misalignment  $\frac{\epsilon_s^y}{\cos\theta_i}$ along the $y$ axis on the IRS. Further projecting  $\frac{\epsilon_s^y}{\cos\theta_i}$ along the $y_2$ axis yields misalignment $\frac{\epsilon_s^y\cos\theta_r}{\cos\theta_i}$, which in turn causes the overall misalignment $\frac{\epsilon_s^y\cos\theta_r}{\cos\theta_{rl}\cos\theta_i}$ along the $y_3$ axis on the lens. Furthermore, moving the IRS by $\epsilon_r^z$ along the $z$ axis is equivalent to keeping the IRS fixed but having a misalignment of $-\epsilon_r^z(\tan\theta_i+\tan\theta_r)$ along the $y$ axis on the IRS. Projecting this misalignment first along the $y_2$ axis and then along the $y_3$ axis leads to $\frac{-\epsilon_r^z(\tan\theta_i+\tan\theta_r)\cos\theta_r}{\cos\theta_{rl}}=\frac{\epsilon_r^z\sin(\theta_i+\theta_r)}{\cos\theta_{rl}\cos\theta_i}$. This completes the proof.  
\end{IEEEproof}

Fluctuations $(\epsilon_s^y,\epsilon_r^z,\epsilon_l^y)$ are typically modelled as RVs. A widely-accepted model for building sway assumes independent zero-mean Gaussian fluctuations \cite{Steve_pointing_error,TCOM_2020}, i.e., $\epsilon_i^{j}\sim\mathcal{N}(0,\sigma^2_i),$ $\,i\in\{s,r,l\},$ $\,j\in\{y,z\}$, where $\sigma_i^2$ denotes the variance of $\epsilon_i^j$. Therefore, from \eqref{Eq:Misalignment}, the misalignment also follows a zero-mean Gaussian distribution, i.e., $u\sim\mathcal{N}\big(0,\sigma^2_u\big)$ with variance 
\begin{IEEEeqnarray}{lll}\label{Eq:varu_2D}
\sigma^2_u=\frac{1}{\cos^2\theta_{rl}}\Big(\frac{\cos^2\theta_r}{\cos^2\theta_i}\sigma_s^2+\frac{\sin^2(\theta_i+\theta_r)}{\cos^2\theta_i}\sigma_r^2+\sigma_l^2\Big).
\end{IEEEeqnarray}

\subsection{PDF of GML}
In order to obtain a statistical channel model for the GML $h_g$, first the power collected by the PD (i.e., the conditional GML) has to be derived as a function of $u$. To do so, we can use the exact expressions in \eqref{Eq:Pt} and replace $(\rho_1,\rho_2)$ with $(|u-a_l|,u+a_l)$, assuming that the IRS is sufficiently large such that the lens is located in region $\mathcal{R}$ defined before Lemma~\ref{Lem:Truncated Gauss}. However, the resulting expressions are rather complicated and do not provide useful insights. Thus, to provide insights, we approximate $h_g$ as a function of $u$ as follows
\begin{IEEEeqnarray}{lll} \label{Eq:A0Approx}
h_g\approx A_0 \exp\Big(\frac{-2u^2}{tw^2(d_{e2e},\hat{w}_0)}\Big),
\end{IEEEeqnarray}
where $A_0=\mathrm{erf}\left(\nu\right)$, $t=\frac{\sqrt{\pi}\mathrm{erf}\left(\nu\right)}{2\nu\exp(-\nu^2)\cos^2\theta_{rl}}$, and $\nu=\frac{\sqrt{2}\cos\theta_{rl}a_l}{w(d_{e2e},\hat{w}_0)}$.
The derivation of \eqref{Eq:A0Approx} follows the steps in \cite[Appendix]{Steve_pointing_error} and for completeness and due to space constraints is provided in \cite[Appendix~C]{arXivIRS2020} which is an extended version of this paper. 
 We verify the accuracy of \eqref{Eq:A0Approx} in Fig.~\ref{Fig:CondGML} in Section~VII. Using this approximation, the PDF of $h_g$ is given in the following proposition.


\begin{prop}\label{Prop:2} Assuming Gaussian fluctuations for the building sway, the PDF of $h_g$ is given~by
\begin{IEEEeqnarray}{lll} \label{Eq:PDFs}
f_{h_g}(h_g)=\frac{\sqrt{\varpi}}{2A_0\sqrt{\pi}}\left[\ln\Big(\frac{A_0}{h_g}\Big)\right]^{-\frac{1}{2}}\left(\frac{h_g}{A_0}\right)^{\varpi-1},\quad 0\leq h_g\leq A_0,\qquad
\end{IEEEeqnarray}
where $\varpi=\frac{tw^2(d_{e2e},\hat{w}_0)}{4\sigma^2_u}$.
\end{prop}
\begin{IEEEproof}
Eq. \eqref{Eq:PDFs} can be obtained by exploiting the relation between $u$ and $h_g$ in \eqref{Eq:A0Approx} and the fact that $u$ follows a zero-mean Gaussian distribution.
\end{IEEEproof}


\section{Extension to 3D System Model}
Before studying the 3D system, let us first recall the following insights from the analysis of the 2D system, cf. Sections~III and IV. 
\textit{i)} The concept of an equivalent mirror-assisted system enabled us to analyze the original IRS-assisted system using the image method from geometric optics. 
\textit{ii)} The effect of beam truncation can be neglected if the IRS is sufficiently large. \textit{iii)} The statistical analysis can be significantly simplified by considering only the components of the Tx, IRS, and Rx fluctuation  variables that have a non-negligible impact on the GML. Exploiting these insights, we now extend the models developed so far to the 3D case.

\subsection{Conditional GML}

In the following, we first find the power density distribution across the IRS. Subsequently, similar to the 2D case, we develop an equivalent  mirror-assisted system for the considered 3D system based on which we derive the fraction of  power collected by the PD at the Rx (i.e., the conditional GML). To simplify the analysis, we assume that the IRS size is sufficiently large to neglect  beam truncation. 

\subsubsection{Power Density Distribution}
The following lemma presents the power density distribution across the IRS.

\begin{lem}\label{lem:PowerDistribution3D} 
The power density distribution of the Gaussian beam originating from the LS at point $\mathbf{a}=(x,y)^{\mathsf{T}}$ on the IRS is obtained as
	\begin{IEEEeqnarray}{lll} \label{Eq:Iirslem3D}
		I^{\mathrm{irs}}_{\mathrm{G}}(x,y|d_{sr},\mathbf{\Psi}_i,w_0)=\frac{2\cos\theta_i}{\pi w^2(d_{sr},w_0)}\exp(-2\mathbf{a}^{\mathsf{T}}\mathbf{S}_{w_0}^{d_{sr}}\mathbf{T}_{\theta_i}^2\mathbf{a}),
	\end{IEEEeqnarray}
	where $\mathbf{S}_{w_0}^{d_{sr}}=\frac{1}{w^2(d_{sr},w_0)}\mathbf{I}_2$.
\end{lem}
\begin{IEEEproof}
Please refer to Appendix~\ref{App:powdens_AG}.
\end{IEEEproof}

Note that the contours of the laser beam power density distribution across the IRS are ellipsoids whose major and minor axes are the main diagonal elements of $(2\mathbf{S}_{w_0}^{d_{sr}}\mathbf{T}_{\theta_i}^2)^{-1/2}$.

\subsubsection{Equivalent Mirror-assisted System}
Similar to the 2D case, the equivalent mirror-assisted system is constructed as follows: \textit{i)} The IRS is replaced by a mirror. \textit{ii)} The LS is replaced by a new LS, denoted by LS$_\mathrm{n}$, which is located in a different position such that it has distance $d_{sr}$ from the mirror and its beam hits the mirror with an incident angle equal to the reflection angle in the original IRS-assisted system, i.e., $\mathbf{\Psi}_r=(\theta_r,\phi_r)$. Therefore, from geometric optics, the angle of reflection at the mirror is $\mathbf{\Psi}_r$. \textit{iii)} Unlike the 2D case, in the 3D system, a simple Gaussian beam originating from the LS$_\mathrm{n}$ cannot provide the same power density  distribution across the mirror as that given in \eqref{Eq:Iirslem3D} for any desired reflection angle $\mathbf{\Psi}_r=(\theta_r,\phi_r)$. However, a \textit{rotated astigmatic} Gaussian beam can generate the power density distribution in \eqref{Eq:Iirslem3D} for any $\mathbf{\Psi}_r$. Assuming the same $\hat{x}\hat{y}\hat{z}$-coordinate system as in Section II.B, an \textit{un-rotated} astigmatic Gaussian beam (also referred to as \textit{orthogonal} or \textit{simple} astigmatic Gaussian beam) propagating in $+\hat{z}$ direction has the following electric field  \cite{Kochkina2013}
\begin{IEEEeqnarray}{lll}\label{Eq:Eastigmatic}
	E(\hat{x},\hat{y},\hat{z})=&E_0\exp\bigg(-\Big(\frac{\hat{x}^2}{w^2(\hat{z},w_{0_1})}+\frac{\hat{y}^2}{w^2(\hat{z},w_{0_2})}\Big)\bigg)\exp\big(j\phi_{\mathrm{AG}}(\hat{\mathbf{a}},\hat{z},\mathbf{w}_0)\big),\quad\text{with phase}\quad
\end{IEEEeqnarray}
\begin{IEEEeqnarray}{lll}\label{Eq:Phase_Eastigmatic}
	\phi_{\mathrm{AG}}(\hat{\mathbf{a}},\hat{z},\mathbf{w}_0)=-k\hat{z}-k\Big(\frac{\hat{x}^2}{2R(\hat{z},w_{0_1})}+\frac{\hat{y}^2}{2R(\hat{z},w_{0_2})}\Big)+\psi(\hat{z},w_{0_1},w_{0_2}),
\end{IEEEeqnarray}
where  $w_{0_1}$ and $w_{0_2}$ are the waist radii of the beam and $\mathbf{w}_{0}=(w_{0_1},w_{0_2})$. Moreover,  $w(\cdot,\cdot)$ and $R(\cdot,\cdot)$ are given in Section II.B, $\hat{\mathbf{a}}=(\hat{x},\hat{y})^{\mathsf{T}}$, $\psi(\hat{z},w_{0_1},w_{0_2})=\frac{1}{2}\big(\tan^{-1}(\frac{\hat{z}}{z_{R_1}})$ $+\tan^{-1}(\frac{\hat{z}}{z_{R_2}})\big)$, and $z_{R_i}=\frac{\pi w^2_{0_i}}{\lambda}$. A \textit{rotated astigmatic} Gaussian beam (also referred to as \textit{general} astigmatic Gaussian beam) is obtained from \eqref{Eq:Eastigmatic} and \eqref{Eq:Phase_Eastigmatic} by substituting $\hat{\mathbf{a}}$ by $\mathbf{R}_{\varphi}\hat{\mathbf{a}}$, where $\varphi$ is the angle by which the beam footprint is rotated counter-clockwise. Therefore, the power density of a general astigmatic Gaussian beam across a plane perpendicular to the beam direction, i.e., $+\hat{z}$, is given~by
\begin{IEEEeqnarray}{lll}\label{Eq:Iastigmatic}
	I^{\mathrm{orth}}_{\mathrm{AG}}(\hat{x},\hat{y}|\hat{z},\mathbf{w}_0,\varphi)=I^{\max}_{\mathrm{AG}}(\hat{z},\mathbf{w}_0)\exp\big(-2\hat{\mathbf{a}}^{\mathsf{T}}\mathbf{R}_{\varphi}^{\mathsf{T}}\mathbf{S}_{\mathbf{w}_0}^{\hat{z}}\mathbf{R}_{\varphi}\hat{\mathbf{a}}\big),
\end{IEEEeqnarray}
where $I^{\max}_{\mathrm{AG}}(\hat{z},\mathbf{w}_0)=\frac{2}{\pi w(\hat{z},w_{0_1})w(\hat{z},w_{0_2})}$ and $\mathbf{S}_{\mathbf{w}_0}^{\hat{z}}=\mathrm{diag}\big\{\frac{1}{w^2(\hat{z},w_{0_1})} ,\frac{1}{w^2(\hat{z},w_{0_2})}\big\}$.

 \begin{corol}\label{corol:equivalence3D} 
 Let us consider a new LS whose location is specified by $\hat{\boldsymbol{\Psi}}_i=(\hat{\theta}_i,\hat{\phi}_i)$ and $d_{sr}$ and  which emits a general astigmatic Gaussian beam with waist radius parameter $\hat{\mathbf{w}}_{0}=(\hat{w}_{0_1},\hat{w}_{0_2})$ and rotation angle $\hat{\varphi}$. Moreover, let $\lambda_i$ and $\boldsymbol{\nu}_i$ denote the $i$-th eigenvalue and the corresponding eigenvector of matrix $\mathbf{T}_{\hat{\theta}_i}^{\mathsf{-T}}\mathbf{R}_{\hat{\phi}_i}^{\mathsf{T}}\mathbf{S}_{w_0}^{d_{sr}}\mathbf{T}_{\theta_i}^2\mathbf{R}_{\hat{\phi}_i}\mathbf{T}_{\hat{\theta}_i}^{-1}$, respectively. For  \textit{any given} $\hat{\boldsymbol{\Psi}}_i$, the power density distribution across the IRS caused by the new LS becomes identical to that created by the original LS, i.e.,    
	\begin{IEEEeqnarray}{lll}\label{Eq:I_irs_AG_I_irs_G}
		I^{\mathrm{irs}}_{\mathrm{AG}}(x,y|d_{sr},\hat{\boldsymbol{\Psi}}_i,\hat{\mathbf{w}}_{0},\hat{\varphi}) =I^{\mathrm{irs}}_{\mathrm{G}}(x,y|d_{sr},\boldsymbol{\Psi}_i,w_0),
	\end{IEEEeqnarray}
	if $\hat{\mathbf{w}}_{0}=(\hat{w}_{0_1},\hat{w}_{0_2})$ and $\hat{\varphi}$ are chosen as the solutions to equations $w(d_{sr},\hat{w}_{0_i})=\frac{1}{\sqrt{\lambda_i}},i=1,2,$ and $\mathbf{R}_{\hat{\varphi}}=[\boldsymbol{\nu}_1,\boldsymbol{\nu}_2]^{\mathsf{T}}$, respectively.   
\end{corol}

\begin{IEEEproof}
Please refer to Appendix~\ref{App:equivalence3D}. 
\end{IEEEproof}

The above corollary states that the power density distribution across the IRS generated by the LS is equivalent to the power density distribution generated by a new LS with \textit{arbitrary} incident angle but with general astigmatic Gaussian beam with waist radii $\hat{\mathbf{w}}_{0}$ and rotation angle $\hat{\varphi}$ whose values are given in Corollary~\ref{corol:equivalence3D}. Therefore, according to Corollary~\ref{corol:equivalence3D} for $\hat{\mathbf{\Psi}}_i=\mathbf{\Psi}_r$, the power density distributions across the IRS and the mirror in the equivalent system are identical.

\subsubsection{Phase-shift Design} The location of LS$_\mathrm{n}$ and the properties of its emitted beam (i.e., $\hat{\mathbf{w}}_0$ and $\hat{\varphi}$) determine the phase distribution of the wave reflected by the mirror. Now, we design the phase-shift profile, $\Delta\phi(x,y|\mathbf{\Psi}_r,\mathbf{\Psi}_i,w_0)$, across the IRS such that the phases of the reflected beams in both systems are the same. This leads to the same reflected electric fields in both systems. Therefore, the angle of reflection in both systems is also identical and is given by $\mathbf{\Psi}_r$.

\textit{Proposed Phase-shift Profile:}
The phase-shift profile that is needed across the IRS to produce an angle of reflection $\mathbf{\Psi}_r$ for an angle of incident $\mathbf{\Psi}_i$ is~given~by 
\begin{align} \label{Eq:betaxMain_3D}
\Delta\phi(x,y|\mathbf{\Psi}_r,\mathbf{\Psi}_i,w_0)=&\pi+\phi_{\mathrm{AG}}(\mathbf{R}_{\hat{\varphi}}\mathbf{T}_{\theta_r}\mathbf{R}_{-\phi_r}\mathbf{a},d_{sr}+x\sin\theta_r\cos\phi_r +y\sin\theta_r\sin\phi_r,\hat{\mathbf{w}}_0)\nonumber\\
&-\phi_{\mathrm{G}}(\sqrt{\mathbf{a}^{\mathsf{T}}\mathbf{T}^2_{\theta_i}\mathbf{a}},d_{sr}+x\sin\theta_i,{w}_0),\quad
\end{align}
where  $\phi_{\mathrm{G}}(\cdot,\cdot,\cdot)$ is the phase of the Gaussian beam originating from the LS, which is given in \eqref{Eq:phase_E}, and $\phi_{\mathrm{AG}}(\cdot,\cdot,\cdot)$ is the phase of the general astigmatic Gaussian beam originating from LS$_{\mathrm{n}}$, which is given in \eqref{Eq:Phase_Eastigmatic}. The proposed phase-shift profile in \eqref{Eq:betaxMain_3D} reduces to the well-known constant phase-shift gradient design, i.e., $\Delta\phi(x,y|\mathbf{\Psi}_r,\mathbf{\Psi}_i,w_0)=\pi+k\big(x(\sin\theta_i-\sin\theta_r\cos\phi_r)-y\sin\theta_r\sin\phi_r\big)$ \cite{GSnellLaw}, if $d_{sr}\gg \max\Big\{\frac{\pi w_0^2}{\lambda},$ $\frac{|\mathbf{a}^{\mathsf{T}}(\mathbf{T}^2_{\theta_i}-\mathbf{R}_{\phi_r}\mathbf{T}^2_{\theta_r}\mathbf{R}_{-\phi_r})\mathbf{a}|}{\lambda}\Big\}$ holds.
 In the following, similar to the 2D case, we use the image method from  geometric optics for the equivalent system to obtain the power collected by~the~PD.

\subsubsection{Conditional GML}
Assuming a sufficiently large mirror in the equivalent system, LS$_{\mathrm{n}}$ can be mirrored w.r.t. the mirror plane, creating a virtual LS denoted by VS$_{\mathrm{n}}$ which emits the same beam as LS$_{\mathrm{n}}$, cf. Corollary~\ref{corol:equivalence3D}, and which has distance $d_{e2e}$ from the PD. Then, as shown in Appendix~\ref{App:GML_3D}, the fraction of power collected by the PD (i.e., the conditional GML) is given as follows 
\begin{IEEEeqnarray}{lll} \label{Eq:hg_3D}
h_g(u) \approx A_0 \exp\Big(-\frac{2\|\mathbf{u}\|^2}{t}\Big),
\end{IEEEeqnarray}
where $t=\frac{\pi a_l^2}{4\nu_1\nu_2}\sqrt{\frac{\pi\mathrm{erf}(\nu_1)\mathrm{erf}(\nu_1)}{\nu_1\nu_2\exp(-(\nu_1^2+\nu_2^2))}}$,  $\nu_{1}=a_l\sqrt{\frac{\pi\delta_1}{2}}$, and  $\nu_{2}=a_l\sqrt{\frac{\pi\delta_2}{2}}$. Here, $\delta_i,\,i=1,2,$ are the eigenvalues of matrix $\mathbf{T}_{\theta_{rl}}^{\mathsf{T}}\mathbf{R}_{\hat{\varphi}}^{\mathsf{T}}\mathbf{S}_{\hat{\mathbf{w}}_0}^{d_{e2e}}\mathbf{R}_{\hat{\varphi}}\mathbf{T}_{\theta_{rl}}$, where  $\mathbf{S}_{\hat{\mathbf{w}}_0}^{d_{e2e}}=\mathrm{diag}\big\{\frac{1}{w^2(d_{e2e},\hat{w}_{0_1})},\frac{1}{w^2(d_{e2e},\hat{w}_{0_2})}\big\}$. Moreover, $\mathbf{u}$ denotes the vector of misalignment between the beam footprint center and the center of the lens and $A_0$ denotes the maximum fraction of optical power captured by the lens at $\|\mathbf{u}\|=0$ and is given by $A_0=\frac{\pi^2 \cos\theta_{rl}I^{\max}_{\mathrm{AG}}(d_{e2e},\hat{\mathbf{w}}_0) a_l^2}{4\nu_1\nu_2}\mathrm{erf}(\nu_{1})\mathrm{erf}(\nu_{2})$. In the following, we derive a statistical GML model based on \eqref{Eq:hg_3D} incorporating the impact of the IRS position fluctuations due to building sway.

\begin{figure*}[t]
	\centering
	\includegraphics[valign=c,width=1\linewidth]{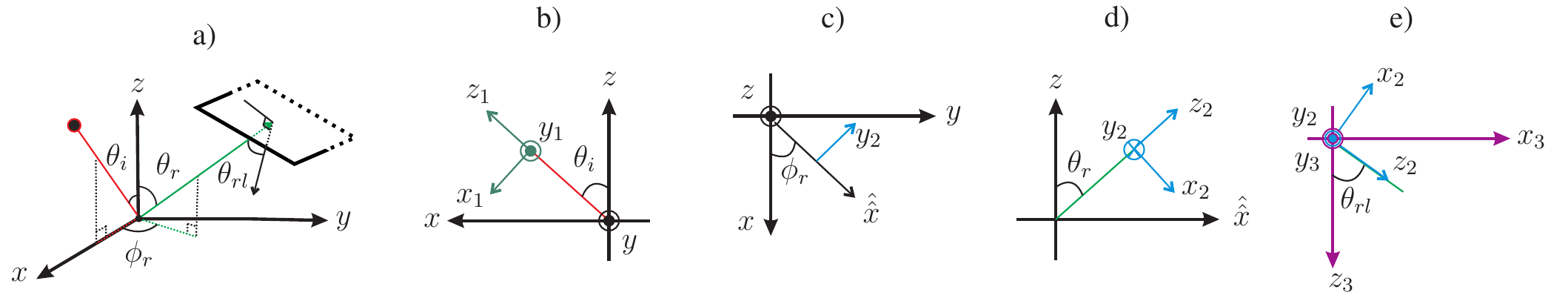}
	\caption{Illustration of 3D coordinate systems adopted to characterize the LS, IRS, and lens fluctuations as well as the misalignment across the lens plane. a) 3D illustration of the positions of LS, IRS, and lens. b) LS coordinate system, $x_1y_1z_1$, w.r.t. IRS coordinate system, $xyz$. c) IRS coordinate system, $xyz$, w.r.t.  coordinate system for the  plane perpendicular to the reflected beam, $x_2y_2z_2$, in the $x-y$ plane. Here, $\hat{\hat{x}}$ is the axis perpendicular to the $y_2$ axis in the IRS plane. d) $xyz$-coordinate system w.r.t. $x_2y_2z_2$-coordinate system in the $\hat{\hat{x}}-z$ plane. e) $x_2y_2z_2$-coordinate system  w.r.t. the lens coordinate system, $x_3y_3z_3$.
	 The same color code is used as in Fig.~\ref{Fig:Coordinate_Sys} for the 2D case.}
	\label{Fig:CooSys3D}
\end{figure*}

\subsection{Statistical GML Model}
Similar to the statistical analysis for 2D systems in Section~IV,  we assume Gaussian fluctuations due to building sway for the LS, IRS, and Rx lens. To this end, we define suitable right-handed coordinate systems to formally characterize the fluctuations of LS, IRS, and lens and the overall misalignment across the lens, cf. Fig.~\ref{Fig:CooSys3D}.

\textbf{LS:}  For the LS, fluctuations along the $z_1$ axis have negligible impact on $h_g$; hence, to characterize the fluctuations of the LS position, we need only two variables along the $x_1$ and $y_1$ directions, denoted by $\epsilon_s^{x},\,\epsilon_s^{y}\sim\mathcal{N}(0,\sigma_s^2)$, respectively,  cf. Fig.~\ref{Fig:CooSys3D}b).

\textbf{IRS:} Since we assume a sufficiently large IRS such that beam truncation can be neglected, the fluctuations of the IRS within its plane, i.e., the $x-y$ plane, can be neglected. Therefore, only the fluctuations along  the $z$ axis, denoted by $\epsilon_r^{z}\sim\mathcal{N}(0,\sigma_r^2)$, are considered,  cf. Fig.~\ref{Fig:CooSys3D}c) and d). 

\textbf{Lens:} Similar to the LS, the fluctuations along the reflected beam direction, $z_2$,  can be neglected. Let $\epsilon_l^{x},\,\epsilon_l^{y}\sim\mathcal{N}(0,\sigma_l^2)$ denote the fluctuations along the $x_2$ and $y_2$ axes, respectively,  cf. Fig.~\ref{Fig:CooSys3D}e). 

The misalignment vector $\mathbf{u}$ defined in the lens plane, i.e., in the $x_3-y_3$ plane, is provided in  the following lemma.

\begin{lem}\label{Lem:Misalignment3D} 
	The misalignment vector $\mathbf{u}$  as a function of $\boldsymbol{\epsilon}_s^{xy}=(\epsilon_s^{x},\epsilon_s^{y})^{\mathsf{T}}$, $\epsilon_r^{z}$, and $\boldsymbol{\epsilon}_l^{xy}=(\epsilon_l^{x},\epsilon_l^{y})^{\mathsf{T}}$ is~obtained~as~follows
	\begin{IEEEeqnarray}{lll} \label{Eq:Misalignment3D}
		\mathbf{u}=(u_1,u_2)^{\mathsf{T}}=\mathbf{T}_{\theta_{rl}}^{-1}(\mathbf{T}_{\theta_r}\mathbf{R}_{-\phi_r}\mathbf{T}_{\theta_i}^{-1}\boldsymbol{\epsilon}_s^{xy}-\mathbf{T}_{\theta_r}\mathbf{R}_{-\phi_r}\mathbf{v}\epsilon_r^{z}-\boldsymbol{\epsilon}_l^{xy}),
	\end{IEEEeqnarray}
	where $\mathbf{v}=(\tan\theta_r\cos\phi_r-\tan\theta_i,\,\tan\theta_r\sin\phi_r)^{\mathsf{T}}$.
\end{lem}
\begin{IEEEproof}
The proof follows from projecting the LS, IRS, and lens position fluctuations along the lens plane, i.e., in the $x_3$ and $y_3$ directions. To this end, we use the following result. Let us consider two coordinate systems $\hat{x}\hat{y}\hat{z}$  and $\tilde{x}\tilde{y}\tilde{z}$, where the latter system is transformed to the former by first counter-clockwise rotation of $\tilde{x}$ and $\tilde{y}$ around the $\tilde{z}$ axis by angle $\phi$ and then counter-clockwise rotation of  the resulting new $\tilde{x}$ axis as well as the $\tilde{z}$ axis around the new $\tilde{y}$ axis by angle $\theta$. Therefore, $\theta$ is the angle between the $\hat{z}$ and $\tilde{z}$ axes and between the $\hat{x}$ and $\tilde{x}$ axes and $\phi$ is the angle between the $\hat{y}$ and $\tilde{y}$ axes. Let us assume that a beam propagates parallel to the $\hat{z}$ axis and let $\hat{\mathbf{a}}$ denote the misalignment in the $\hat{x}-\hat{y}$ plane. Projecting $\hat{\mathbf{a}}$ into non-perpendicular plane $\tilde{x}-\tilde{y}$ yields misalignment $\tilde{\mathbf{a}}$ which is related to  $\hat{\mathbf{a}}$  as $\hat{\mathbf{a}}=\mathbf{T}_{\theta}\mathbf{R}_{-\phi}\tilde{\mathbf{a}}$. For the lens, projecting fluctuation $\boldsymbol{\epsilon}_l^{xy}$  into the $x_3-y_3$ plane yields misalignment $-(\mathbf{T}_{\theta_{rl}}\mathbf{R}_{-\phi_{rl}})^{-1}\boldsymbol{\epsilon}_l^{xy}=-\mathbf{T}_{\theta_{rl}}^{-1}\boldsymbol{\epsilon}_l^{xy}$, where by definition $\phi_{rl}=0$. For the LS, fluctuation $\boldsymbol{\epsilon}_s^{xy}$ is first transformed into the $x-y$ plane, then into the $x_2-y_2$ plane, and finally into the $x_3-y_3$ plane. This leads to misalignment $(\mathbf{T}_{\theta_{rl}}\mathbf{R}_{-\phi_{rl}})^{-1}\mathbf{T}_{2\pi-\theta_r}\mathbf{R}_{-\phi_r}(\mathbf{T}_{\theta_{i}}\mathbf{R}_{-\phi_{i}})^{-1}\boldsymbol{\epsilon}_s^{xy}=\mathbf{T}_{\theta_{rl}}^{-1}\mathbf{T}_{\theta_r}\mathbf{R}_{-\phi_r}\mathbf{T}_{\theta_i}^{-1}\boldsymbol{\epsilon}_s^{xy}$, where we use definitions $\phi_i=\phi_{rl}=0$. Furthermore, moving the IRS with $\epsilon_r^z$ along the $z$ axis is equivalent to keeping the IRS fixed but having a misalignment of $\mathbf{v}\epsilon_r^{z}$ along the $x-y$ plane on the IRS. Projecting this misalignment first into the $x_2-y_2$ plane, and then into the $x_3-y_3$ plane leads to  $(\mathbf{T}_{\theta_{rl}}\mathbf{R}_{-\phi_{rl}})^{-1}\mathbf{T}_{2\pi-\theta_r}\mathbf{R}_{-\phi_r}\mathbf{v}\epsilon_r^{z}=\mathbf{T}_{\theta_{rl}}^{-1}\mathbf{T}_{\theta_r}\mathbf{R}_{-\phi_r}\mathbf{v}\epsilon_r^{z}$. This completes the proof. 
\end{IEEEproof}

From \eqref{Eq:Misalignment3D}, we observe that $\mathbf{u}$ is a zero-mean bivariate Gaussian RV with covariance matrix 
	\begin{IEEEeqnarray}{lll} \label{Eq:CovMatrixU}
	\boldsymbol{\Sigma}=\mathbf{T}_{\theta_{rl}}^{-1}(\mathbf{T}_{\theta_r}\mathbf{R}_{-\phi_r}\mathbf{T}_{\theta_i}^{-1}\mathbf{T}_{\theta_i}^{-\mathsf{T}}\mathbf{R}_{-\phi_r}^{\mathsf{T}}\mathbf{T}_{\theta_r}^{\mathsf{T}}\sigma^2_s +\mathbf{T}_{\theta_r}\mathbf{R}_{-\phi_r}\mathbf{v}\mathbf{v}^{\mathsf{T}}\mathbf{R}_{-\phi_r}^{\mathsf{T}}\mathbf{T}_{\theta_r}^{\mathsf{T}}  \sigma^2_r +\sigma^2_l\mathbf{I}_2)\mathbf{T}_{\theta_{rl}}^{-\mathsf{T}}.
	\end{IEEEeqnarray}
 Therefore, $\|\mathbf{u}\|$ follows a Hoyt (Nakagami-$q$) distribution with mean $\Omega=\chi_1+\chi_2$ and
Nakagami-$q$ fading parameter $q=\Big[\frac{\min\{\chi_1,\chi_2\}}{\max\{\chi_1,\chi_2\}}\Big]^{1/2}$ \cite{Alouini_Pointing,TCOM_2020}, where $\chi_1$ and $\chi_2$ are the eigenvalues of $\boldsymbol{\Sigma}$.
 Exploiting \eqref{Eq:hg_3D} and \eqref{Eq:Misalignment3D}, the PDF of $h_g$ can be obtained as follows
\begin{IEEEeqnarray}{lll} \label{Eq:PDF_h3D}
	f_{h_g}(h_g) =  \frac{\varpi }{A_0}
	\left(\frac{h_g}{A_0}\right)^{\frac{(1+q^2)\varpi}{2q}-1} I_0\left(-\frac{(1-q^2)\varpi}{2q}\ln\left(\frac{h_g}{A_0}\right)\right),
	 \quad 0\leq h_g \leq A_0,
\end{IEEEeqnarray}
where $\varpi = \frac{(1+q^2)t}{4q\Omega}$ is a constant.

\begin{corol}\label{corol:SpecialCase}
For the special case where $\phi_r=\pi$ and $\theta_{rl}=0$, the misalignment $\mathbf{u}$ simplifies to 
	\begin{IEEEeqnarray}{lll} \label{Eq:Misalignment3D_specialcase}
		\mathbf{u}=\Big(\frac{\cos\theta_r}{\cos\theta_i}\epsilon_s^{x}-\frac{\sin(\theta_i+\theta_r)}{\cos\theta_i}\epsilon_r^z -\epsilon_l^{x},\,\epsilon_s^{y}-\epsilon_l^{y}\Big),
	\end{IEEEeqnarray}
	which is again a zero-mean bivariate Gaussian RV. Therefore, the GML follows the distribution given in \eqref{Eq:PDF_h3D} with parameters $q=\Big[\frac{\min\{\sigma^2_{u_{1}},\sigma^2_{u_{2}}\}}{\max\{\sigma^2_{u_{1}},\sigma^2_{u_{2}}\}}\Big]^{1/2}$ and $\Omega=\sigma^2_{u_{1}}+\sigma^2_{u_{2}}$, where $\sigma^2_{u_{1}}=\frac{\cos^2\theta_r}{\cos^2\theta_i}\sigma^2_s+\frac{\sin^2(\theta_i+\theta_r)}{\cos^2\theta_i}\sigma^2_r+\sigma^2_l$ and $\sigma^2_{u_{2}}=\sigma^2_s+\sigma^2_l$. In this case, the parameters in \eqref{Eq:hg_3D} simplify as follows. In particular,
	$A_0=\mathrm{erf}(\nu_1)\mathrm{erf}(\nu_2)$ and $\delta_i=\frac{1}{w^2(d_{e2e},\hat{w}_{0_i})}$ hold, where $\hat{w}_{0_i},\, i=1,2,$ are given as the solutions of $w(d_{sr},\hat{w}_{0_1})=|\frac{\cos\theta_r}{\cos\theta_i}|w(d_{sr},w_0)$ and $w(d_{sr},\hat{w}_{0_2})=w(d_{sr},w_0)$, respectively.  When only the IRS moves, i.e., $\sigma_s=0$ and $\sigma_l=0$ hold, the misalignment becomes a scalar. In this case, the magnitude of the misalignment is given by $|u|=\big|\frac{\sin(\theta_i+\theta_r)}{\cos\theta_i}\epsilon_r^z\big|$ which follows a single-sided Gaussian distribution, i.e., $f_{|u|}(x)=\frac{\sqrt{2}}{\sqrt{\pi\sigma_t^2}}\exp\left(-\frac{x^2}{2\sigma_t^2}\right)$, where $\sigma_t^2=\frac{\sin^2(\theta_i+\theta_r)}{\cos^2\theta_i}\sigma_r^2$. Therefore,  $h_g$  follows the following distribution
\begin{IEEEeqnarray}{lll} \label{Eq:PDF_h3D_specialCase}
	f_{h_g}(h_g) =  \frac{\sqrt{\varrho}}{\sqrt{\pi}A_0}
	\left(\frac{h_g}{A_0}\right)^{\varrho-1} \left[\ln\left(\frac{A_0}{h_g}\right)\right]^{-1/2},
	 \quad 0\leq h_g \leq A_0,
\end{IEEEeqnarray}
	where $\varrho=\frac{t}{4\sigma_t^2}$.
\end{corol}

\begin{remk}\label{Remk:2}
 In the case of a simple mirror instead of an IRS, $\phi_r=\pi$ and $\theta_r=\theta_i$ hold. Therefore, in order to provide a desired angle of reflection, unlike the IRSs, the mirror has to mechanically rotate \cite{Globecom2019}. In this case, the PDF of the GML is the same as in \eqref{Eq:PDF_h3D} with $q=\big[\frac{\sigma^2_s+\sigma^2_l}{\sigma^2_s+\sigma^2_l+4\sigma^2_r\sin^2\theta_i}\big]^{1/2}$ and $\Omega=2\sigma^2_s+2\sigma^2_l+4\sigma^2_r\sin^2\theta_i$ \cite{Globecom2019}. 
\end{remk}

\section{Outage Probability of the 3D Model}
Due to the high data rates of FSO systems, the channel coherence time, which is in the order of a few milliseconds, is much larger than a symbol interval.  Therefore, the outage probability is a relevant metric to evaluate the performance of FSO systems in the presence
of atmospheric turbulence and GML \cite{Safari_AoA}. The outage probability is defined as the probability that the SNR, given by $\gamma=h^2\bar{\gamma}$ with transmit SNR $\bar{\gamma}=\frac{P^2}{\sigma^2_n}$, falls below a predefined threshold, $\gamma_{\mathrm{thr}}$, whose value depends on the adopted transmission scheme and the application of interest. The outage probability can be computed as follows
\begin{IEEEeqnarray}{lll} \label{Eq:outage}
P_{\mathrm{out}}=\mathrm{Pr}\{\gamma\leq\gamma_{\mathrm{thr}}\}=\mathrm{Pr}\{h^2\bar{\gamma}\leq\gamma_{\mathrm{thr}}\}=\int_0^{\sqrt{\gamma_{\mathrm{thr}}/ \bar{\gamma}}}f_h(h)\mathrm{d}h,
\end{IEEEeqnarray}
where $f_h(h)$ denotes the PDF of the composite channel given in \eqref{Eq:channel}. This PDF can be expressed in terms of the following finite integral
\begin{align} 
f_h(h)&=\int_0^{A_0} f_{h_g}\left(h_g\right)f_{h}\left(h|h_g\right)\mathrm{d}h_g \overset{(a)}{=}\int_0^{1}\frac{1}{\eta h_p x}f_{h_g}\left(A_0x\right)f_{h_a}\left(\frac{h}{\eta h_pA_0 x}\right)\mathrm{d}x,\label{Eq:ComFadGen}
\end{align}
where  for equality (a), we used $h=\eta h_p h_g h_a$ in \eqref{Eq:channel} to obtain $f_{h}\left(h|h_g\right)=\frac{1}{\eta h_p h_g}f_{h_a}\left(\frac{h}{\eta h_p h_g}\right)$ and the variable transformation $x=\frac{h_g}{A_0}\in\{0,1\}$. Here, $f_{h_g}(h_g)$ is given in \eqref{Eq:PDF_h3D} and $f_{h_a}(h_a)$ denotes the PDF of the atmospheric turbulence induced fading  which is given by
\begin{IEEEeqnarray}{lll} \label{Eq:LNGG}
f_{h_a}(h_a)=
\begin{cases}
\frac{1}{\sqrt{8\pi\sigma^2}h_a }\exp\left(-\frac{\left(\ln(h_a)+2\sigma^2\right)^2}{8\sigma^2}\right),\quad & \text{LN}\\
\frac{2(\alpha \beta h_a)^{\frac{\alpha+\beta}{2}}}{\Gamma(\alpha)\Gamma(\beta)h_a}K_{\alpha-\beta}\left(2\sqrt{\alpha \beta h_a}\right), & \text{GG}.
\end{cases}
\end{IEEEeqnarray}
In \eqref{Eq:LNGG},  $\sigma^2$ is the variance of the log fading amplitude $\frac{1}{2}\log(h_a)$ and is given by $\sigma^2\approx \frac{\sigma^2_R}{4}$ \cite{Steve_pointing_error,Alouini_Pointing}, where $\sigma_R^2=1.23C_n^2k^{7/6}{d_{e2e}}^{11/6}$ is the Rytov variance, $C_n^2$ is the index of refraction structure parameter, and $\alpha$ and $\beta$ are given as follows \cite{FSO_Survey_Murat}
\begin{IEEEeqnarray}{rll}\label{Eq:alphabeta}
	 \alpha=\left[\exp\bigg(\tfrac{0.49\sigma_R^2}{(1+1.11\sigma_R^{12/5})^{7/6}}\bigg)-1\right]^{-1}\text{and}\,\,\, 
	\beta=\left[\exp\bigg(\tfrac{0.51\sigma_R^2}{(1+0.69\sigma_R^{12/5})^{5/6}}\bigg)-1\right]^{-1}.\quad
\end{IEEEeqnarray}
Note that turbulence conditions can be categorized into two regimes according to the Rytov variance, namely, weak turbulence ($\sigma^2_R<0.3$) which is modelled by the LN distribution and  moderate-to-strong turbulence ($\sigma^2_R\geq 0.3$) which is modelled by the GG distribution \cite{Alouini_Pointing}.  
Unfortunately, the composite fading PDF in \eqref{Eq:ComFadGen} cannot be computed in closed form. However, it involves  only one finite integral that can be easily computed numerically.  
The outage probability for LN and GG turbulence induced fading is given in the following for 3D systems.

\subsubsection{LN Fading}
The outage probability of an IRS-assisted FSO link assuming LN distributed atmospheric turbulence induced fading is obtained by substituting $f_{h_a}(\cdot)$ for LN fading in \eqref{Eq:LNGG} into \eqref{Eq:outage} and \eqref{Eq:ComFadGen}~as~follows
\begin{IEEEeqnarray}{lll} \label{Eq:out_LN}
P_{\mathrm{out}}=&\frac{\varpi}{2}\int_0^{1}x^{\frac{(1+q^2)\varpi}{2q}-1}I_0\Big(-\frac{(1-q^2)\varpi}{2q}\ln\left(x\right)\Big)\bigg[1-\mathrm{erf}\bigg(\frac{0.5\ln\Big(\frac{\bar{\gamma}}{\gamma_{\mathrm{thr}}}\Big)+\ln\Big(\eta h_p A_0x\Big)-2\sigma^2}{\sqrt{8\sigma^2}}\bigg)\bigg]\mathrm{d}x.\quad\,\,\,\,
\end{IEEEeqnarray}

\subsubsection{GG Fading}
Similarly, the outage probability of an IRS-assisted FSO link assuming GG distributed atmospheric turbulence induced fading is obtained as follows
\begin{IEEEeqnarray}{lll} \label{Eq:out_GG}
P_{\mathrm{out}}=\frac{\pi\varpi}{\sin(\pi(\alpha-\beta))}\int_0^{1}x^{\frac{(1+q^2)\varpi}{2q}-1}I_0\Big(-\frac{(1-q^2)\varpi}{2q}\ln\left(x\right)\Big)\big(f(\alpha,\beta,x)-f(\beta,\alpha,x)\big)\mathrm{d}x,\quad\\
\text{with}\quad f(\alpha,\beta,x)=\frac{1}{\Gamma(\alpha)}\left(\frac{\alpha\beta\sqrt{\gamma_{\mathrm{thr}}/\bar{\gamma}}}{\eta h_p A_0}\right)^{\beta}x^{-\beta}{}_1\tilde{F}_2\left(\beta;\beta-\alpha+1,\beta+1;\frac{\alpha\beta\sqrt{\gamma_{\mathrm{thr}}/\bar{\gamma}}}{\eta h_p A_0x}\right),\nonumber
\end{IEEEeqnarray}
where ${}_p\tilde{F}_q\left(a_1,\dots,a_l;b_1,\dots,b_q;z\right)$ denotes the regularized hypergeometric function which is given by ${}_p\tilde{F}_q\big(a_1,$ $\dots,a_l;b_1,\dots,b_q;z\big)=\frac{{}_p F_q\left(a_1,\dots,a_l;b_1,\dots,b_q;z\right)}{\Gamma\left(b_1\right)\dots \Gamma\left(b_q\right)}$ \cite{TableIntegSerie}.


\begin{table}
	\label{Table:Parameter}
	\caption{Simulation Parameters \cite{Safari_AoA,MyC_RAN}.\vspace{-0.3cm}} 
	\begin{center}
		\scalebox{0.65}
		{
			\begin{tabular}{|| c | c  | c | c | c  | c | c | c  ||}
				\hline
				Symbol & $\lambda$ & $\kappa$ & $\eta$ & $\zeta$ & $d_{e2e}$ & $(\theta_i,\phi_i)$ & $(\theta_r,\phi_r)$  \\ \hline
				Definition & wavelength & channel attenuation coefficient & PD responsivity & reflection efficiency & Tx-IRS-Rx distance & incident angles & reflection angles \\ \hline
				Value & $1550$~nm & $0.43\times10^{-3}$~m$^{-1}$ & $0.5$  & $1$ & $900$~m & $(\frac{\pi}{6},0)$ rad & $(\frac{\pi}{5},\frac{7\pi}{8})$ rad    \\ \hline        
			\end{tabular} 
		}
	\end{center}
	
	\begin{center}
		\scalebox{0.65}
		{
			\begin{tabular}{||  c  | c | c | c | c  | c| c  ||}
				\hline
				Symbol & $\theta_{rl}$ & $a_l$ & $a_r$ & $(y_{ls},z_{ls})$ & $(y_r,z_r)$ & $(y_l,z_l)$ \\ \hline
				Definition &  angle between lens plane and reflected beam & lens length/radius & IRS length/dimension & LS location & IRS location & lens location \\ \hline
				Value & $\frac{\pi}{6}$ rad & $2.5$~cm & $10$~cm  & $(-200,346)$~m & $(0,0)$ & $(300,412)$~m     \\ \hline        
			\end{tabular} 
		}
	\end{center}
	\vspace{-0.5cm}
\end{table}

\section{Simulation Results}
Unless stated otherwise, the default values of the parameters used in our simulations are given in Table~1. The simulation results reported in Figs.~\ref{Fig:PDF}, \ref{Fig:Outage_SNR}, and \ref{Fig:Outage_Distance} were obtained based on Monte Carlo simulation and $10^7$ realizations of RVs $\epsilon_i^{j},i\in\{s,r,l\},j\in\{x,y,z\}$.

\begin{figure*}[!tbp]
  \centering
  \begin{minipage}[b]{0.47\textwidth}
  \centering
\includegraphics[valign=c,width=1\linewidth]{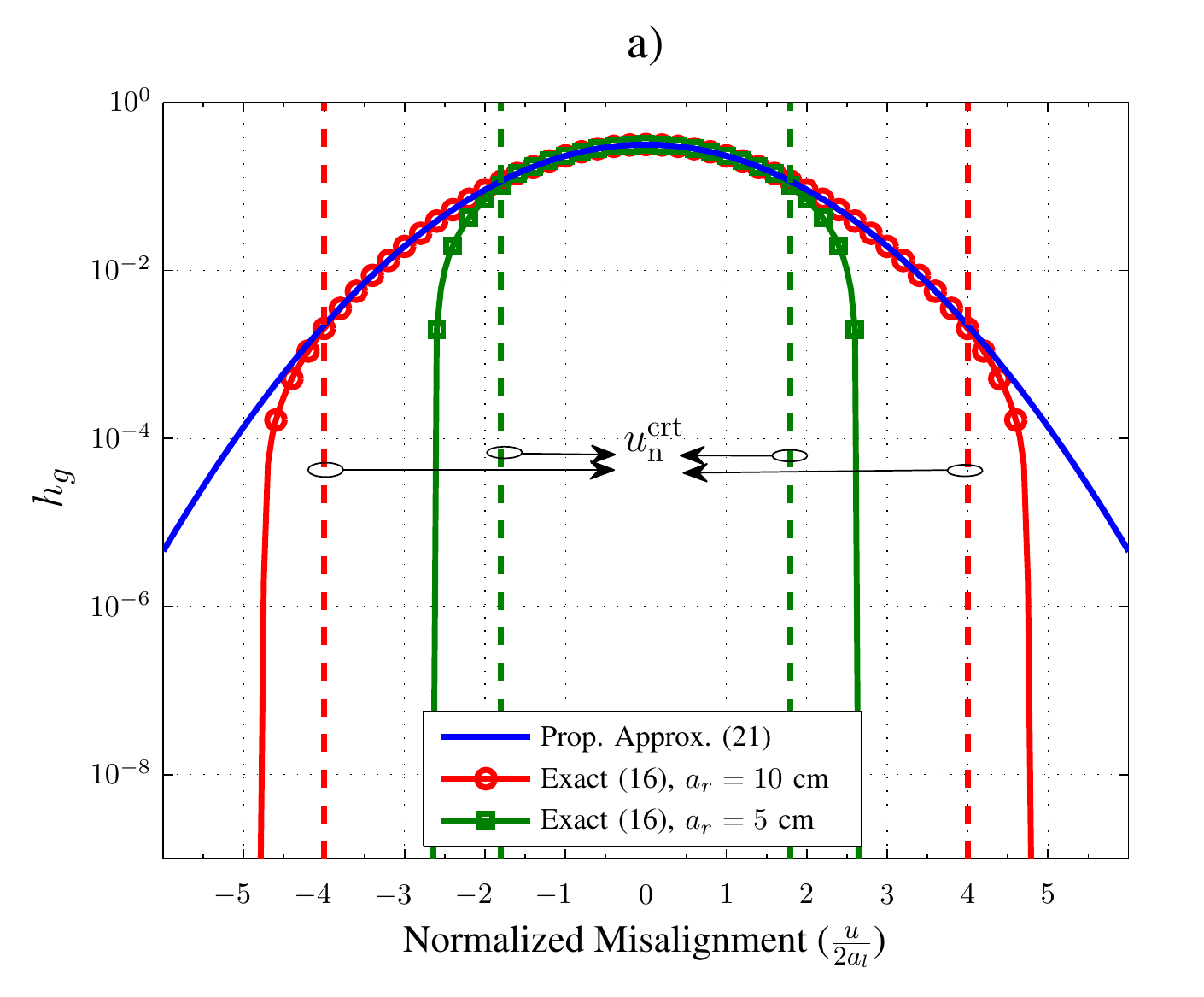}
  \end{minipage}
    \hfill
  \begin{minipage}[b]{0.1\textwidth}
  \end{minipage}
  \hfill
  \begin{minipage}[b]{0.47\textwidth}
  \centering
\includegraphics[valign=c,width=1\linewidth]{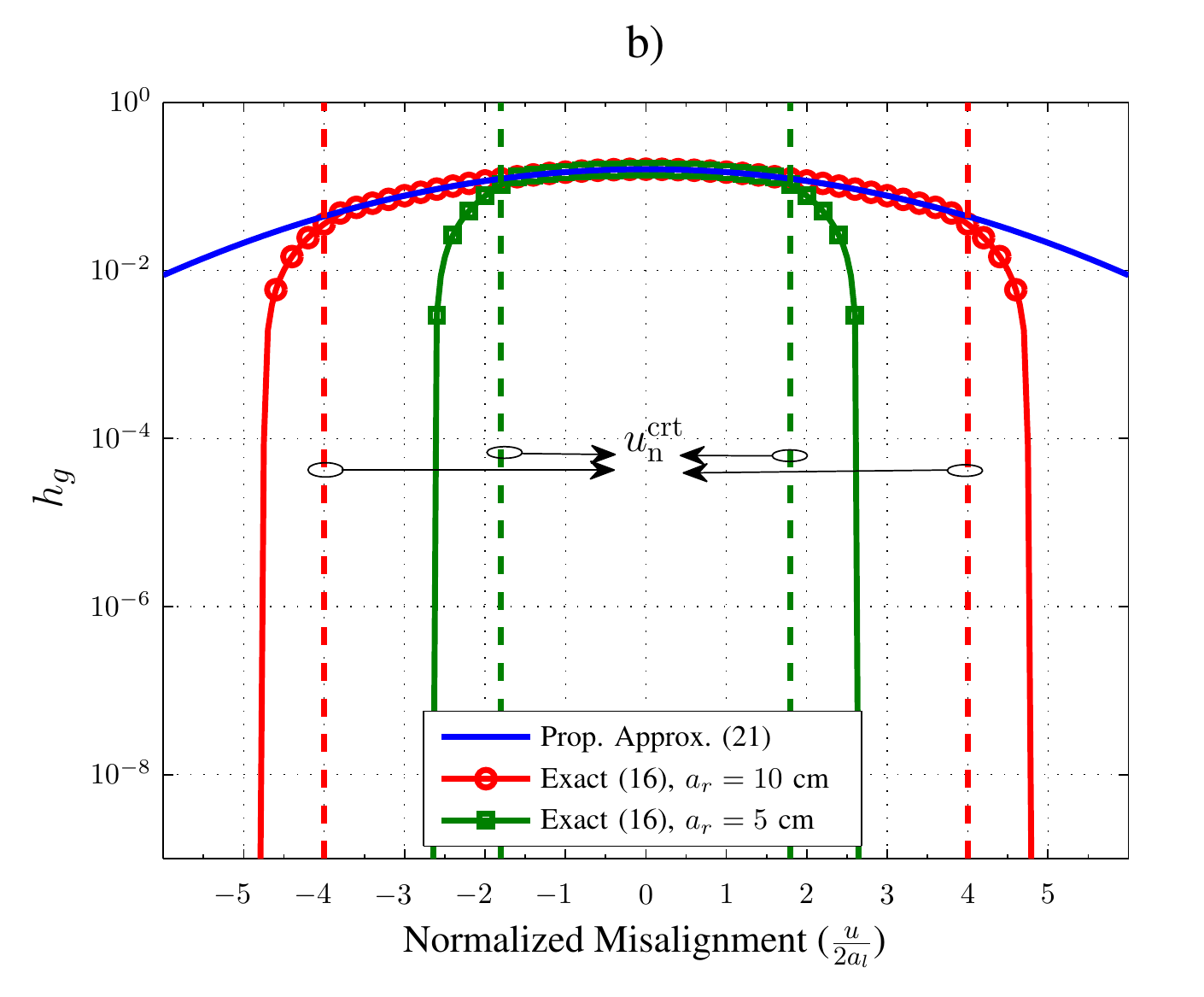}
  \end{minipage}
    \hfill
  \begin{minipage}[b]{0.02\textwidth}
  \end{minipage} \vspace{-3mm}
  \caption{Conditional GML vs. normalized misalignment $u_{\mathrm{n}}=\frac{u}{2a_l}$ for different IRS sizes and a) $\frac{w(d_{e2e},w_0)}{2a_l}=2$ and b)~$\frac{w(d_{e2e},w_0)}{2a_l}=4$. The vertical dashed lines represent critical  values of the normalized misalignment beyond which the proposed approximation in \eqref{Eq:A0Approx} starts to deviate from the exact expression in \eqref{Eq:Pt}.}\vspace{-0.5cm}
  \label{Fig:CondGML}
\end{figure*}

First, in Figs.~\ref{Fig:CondGML}a) and \ref{Fig:CondGML}b), we study the impact of the size of the IRS on the conditional GML in \eqref{Eq:Pt} for two normalized beamwidths at the Rx, namely $\frac{w(d_{e2e},w_0)}{2a_l}\in\{2,4\}$. In particular, we show $h_g$ vs. the normalized misalignment $u_{\mathrm{n}}=\frac{u}{2a_l}$ for IRS lengths of  $a_r=5,10$~cm. As expected, in both figures, we observe that by increasing the magnitude of $u_{\mathrm{n}}$, the channel gain $h_g$ decreases. Moreover, beam truncation occurs if the normalized misalignment exceeds a certain critical value denoted by $u_{\mathrm{n}}^{\mathrm{crt}}$, i.e., when a part of the lens leaves the truncation region $\mathcal{R}$ given before Lemma~\ref{Lem:Truncated Gauss}.
In Fig.~\ref{Fig:CondGML}, we use green and red dashed lines to indicate $u_{\mathrm{n}}^{\mathrm{crt}}$ for $a_r=5$~cm and $a_r=10$~cm, respectively.
Fig.~\ref{Fig:CondGML} shows that the proposed approximation in \eqref{Eq:A0Approx} is accurate when beam truncation does not occur, i.e., for $|u_{\mathrm{n}}|\leq u_{\mathrm{n}}^{\mathrm{crt}}$. However, since the approximation neglects beam truncation, it overestimates $h_g$ when beam truncation does occur, i.e., for $|u_{\mathrm{n}}|> u_{\mathrm{n}}^{\mathrm{crt}}$. Moreover, we observe that, for $a_r=10$~cm, the impact of beam truncation shows itself at larger values of $|u_{\mathrm{n}}|$ compared to $a_r=5$~cm, i.e., $u_{\mathrm{n}}^{\mathrm{crt}}=1.8$ holds for $a_r=5$~cm and $u_{\mathrm{n}}^{\mathrm{crt}}=4$ holds for $a_r=10$~cm. 
 Furthermore, Fig.~\ref{Fig:CondGML} shows that for a reasonable size of the IRS, e.g., $a_r>10$~cm for the considered $d_{e2e}$ ($900$~m) and $a_l$ ($2.5$~cm), the proposed approximation is accurate even for large misalignment magnitudes (e.g., $|u_{\mathrm{n}}|\leq u_{\mathrm{n}}^{\mathrm{crt}}$, where $u_{\mathrm{n}}^{\mathrm{crt}}>4$). Finally, comparing Figs.~\ref{Fig:CondGML}a) with $\frac{w(d_{e2e},w_0)}{2a_l}=2$ and \ref{Fig:CondGML}b) with $\frac{w(d_{e2e},w_0)}{2a_l}=4$, we observe that the maximum value of $h_g$ in  Fig.~\ref{Fig:CondGML}a) is larger than that in Fig.~\ref{Fig:CondGML}b). Moreover, $h_g$ in  Fig.~\ref{Fig:CondGML}a) is more sensitive to misalignment and decays dramatically as the magnitude of the misalignment increases. In other words, in case of no misalignment, it is preferable to have a narrower beam to get a larger $h_g$ and in case of severe misalignment, having a wider beam increases the robustness to the misalignment, although a fraction of power is lost due to beam divergence.

\begin{figure*}[!tbp]
  \centering
  \begin{minipage}[b]{0.47\textwidth}
  \centering
\includegraphics[valign=c,width=1\linewidth]{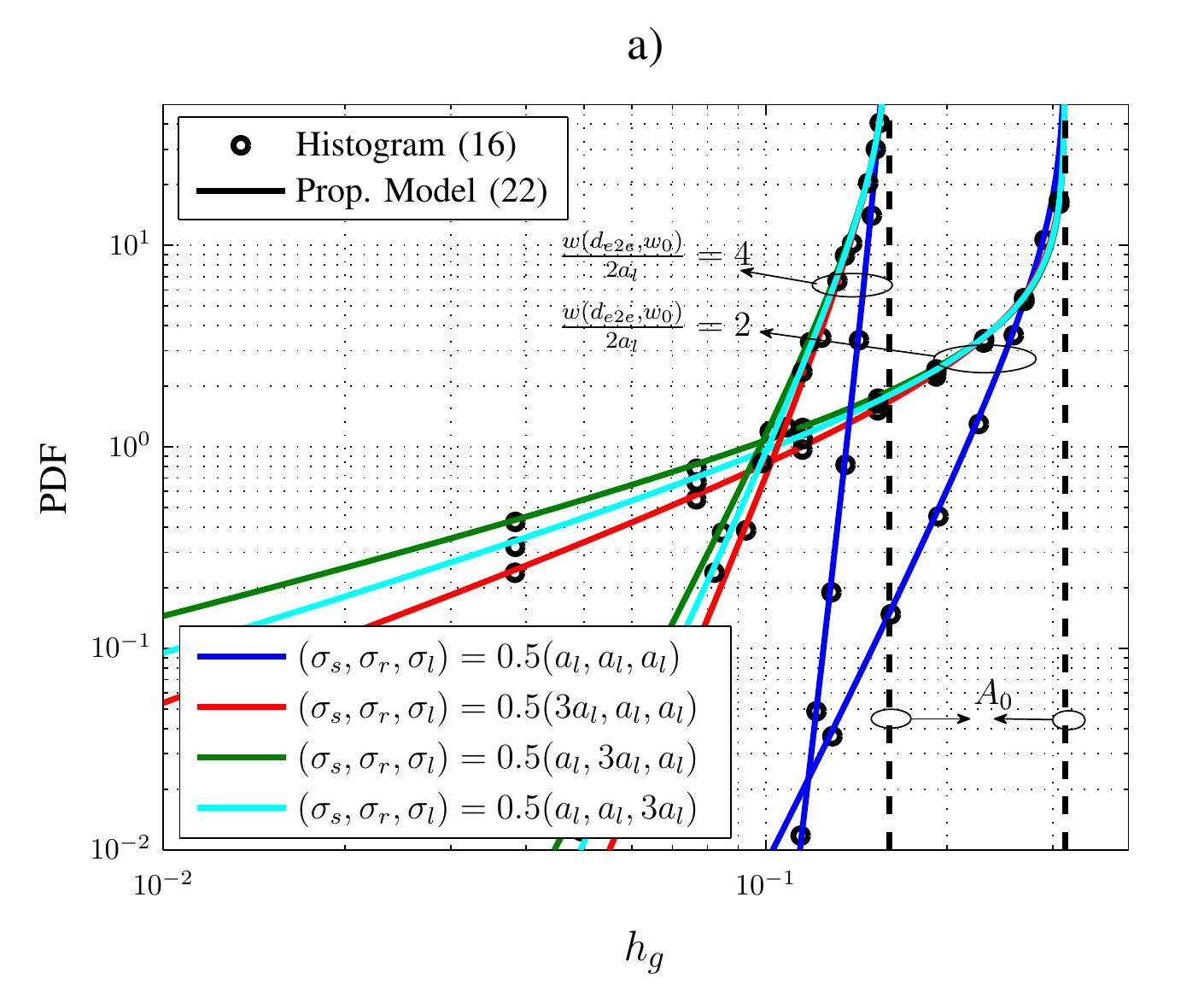}
  \end{minipage}
    \hfill
  \begin{minipage}[b]{0.1\textwidth}
  \end{minipage}
  \hfill
  \begin{minipage}[b]{0.47\textwidth}
  \centering
\includegraphics[valign=c,width=1\linewidth]{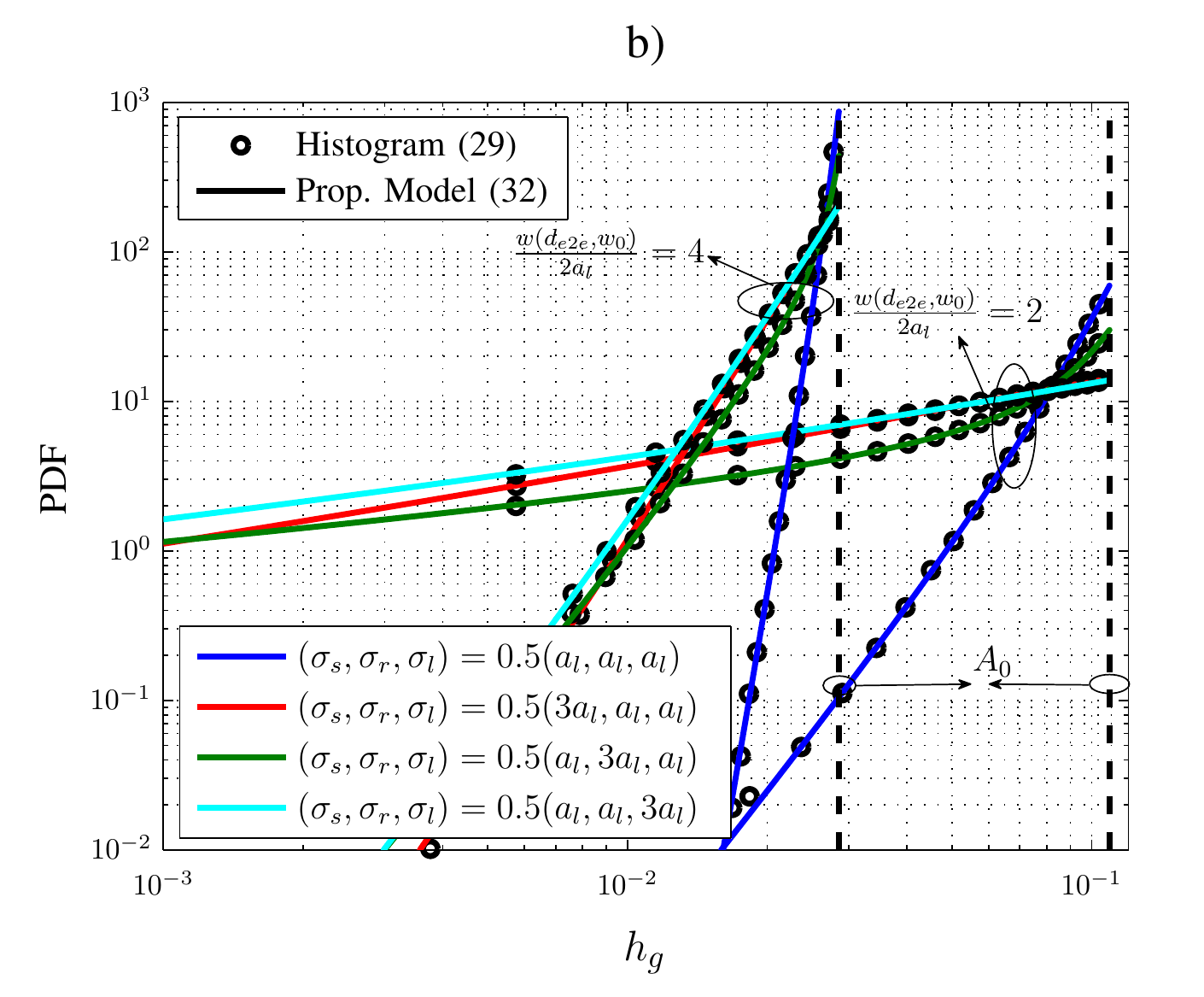}
  \end{minipage}
    \hfill
  \begin{minipage}[b]{0.02\textwidth}
  \end{minipage} \vspace{-3mm}
  \caption{PDF of the GML for  a) 2D and b) 3D system models.}\vspace{-0.5cm}
  \label{Fig:PDF}
\end{figure*}

\begin{figure*}
\centering
\includegraphics[valign=c,width=0.47\linewidth]{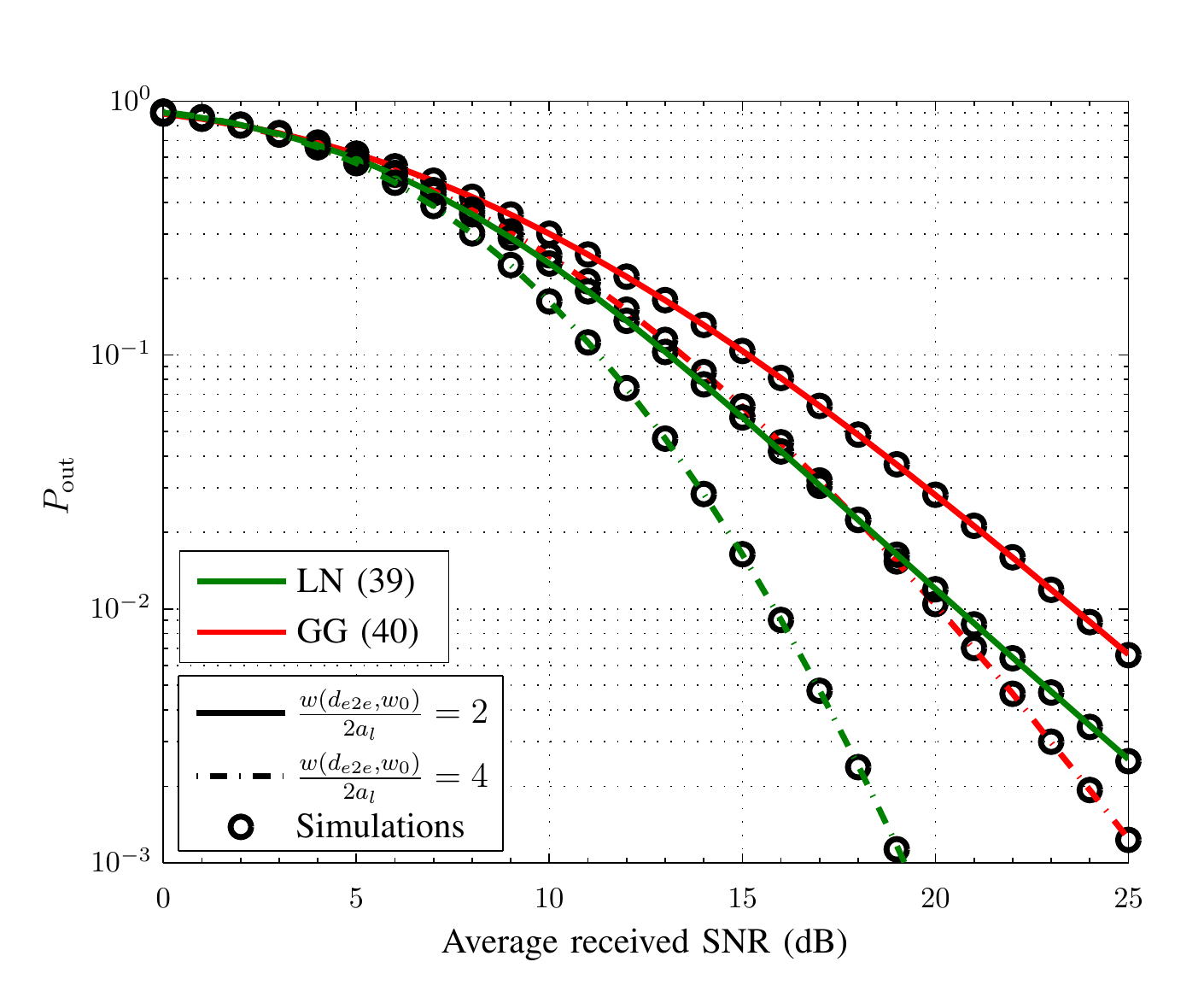}\vspace{-0.3cm}
	\caption{Outage probability vs. average received SNR for $d_{sr}=500$~m, $d_{rl}=600$~m, $\sigma_s=\sigma_r=\sigma_l=0.75a_l$, and $C_n^2=10^{-14}$~m$^{\frac{2}{3}}$ and $C_n^2=1.4\times10^{-14}$~m$^{\frac{2}{3}}$ for LN and GG fading, respectively.}
	\label{Fig:Outage_SNR}
\end{figure*}

Next, in Figs.~\ref{Fig:PDF}a) and \ref{Fig:PDF}b), we study the accuracy of the proposed statistical models for 2D and 3D systems given in \eqref{Eq:PDFs} and \eqref{Eq:PDF_h3D}, respectively. For the simulation results, we plot the histogram of $h_g$ given by \eqref{Eq:Pt} and \eqref{Eq:hg_3D} for 2D and 3D systems, respectively. Fig.~\ref{Fig:PDF} shows the PDF of $h_g$ for two normalized beamwidths, namely $\frac{w(d_{e2e},w_0)}{2a_l}\in\{2,4\}$,  and for four fluctuation scenarios, namely, Scenario~1: $(\sigma_s,\sigma_r,\sigma_l)=0.5(a_l,a_l,a_l)$ where the building sways for the Tx, IRS, and Rx are equal; Scenario~2: $(\sigma_s,\sigma_r,\sigma_l)=0.5(3a_l,a_l,a_l)$ where the building sway for the Tx is more severe than that for the IRS and Rx; Scenario~3: $(\sigma_s,\sigma_r,\sigma_l)=0.5(a_l,3a_l,a_l)$ where the building sway for the IRS is more severe than that for the Tx and Rx; Scenario~4: $(\sigma_s,\sigma_r,\sigma_l)=0.5(a_l,a_l,3a_l)$ where the building sway for the Rx is more severe than that for the Tx and IRS. First, we note that Fig.~\ref{Fig:PDF} shows an excellent agreement between the proposed analytical statistical models and the simulation results for both the 2D and 3D systems. This is due to the fact that the impact of beam truncation, which has been neglected in the analysis, is negligible as it occurs with small probability for the considered parameter values. Comparing Scenarios 2-4 with Scenario 1 shows that, for a given beamwidth, increasing the variance of the fluctuations deteriorates the channel quality, i.e., the probability of having smaller channel gains increases and the PDF becomes more heavily tailed. Moreover, we observe from Fig.~\ref{Fig:PDF}a) that, for the 2D case and a given beamwidth, the building sway of the IRS has the largest impact and building sway of the Tx has the smallest impact on $h_g$. On the other hand, for the 3D case and a given beamwidth, the building sway of the Rx has the largest impact and building sway of the IRS has the smallest impact on $h_g$ in Fig.~\ref{Fig:PDF}b). The different behaviors in the 2D and 3D cases can be understood by investigating the statistics of their corresponding misalignments $u$ and $\mathbf{u}$, respectively.
In particular, for the 2D case, the standard deviation (SD) of $u$ is given by $\sigma_u=1.08\sigma_s+1.22\sigma_r+1.15\sigma_l$, cf. \eqref{Eq:varu_2D}, and
for the 3D case, the SDs of the components of $\mathbf{u}$ are given by $(\sigma_{u_1},\sigma_{u_2})=(1.05,1.02)\sigma_s+(1.17,0.2)\sigma_r+(1.15,1)\sigma_l$, cf. \eqref{Eq:Misalignment3D}. Therefore, for the 2D case, $\sigma_r$ has the largest impact on the misalignment, whereas for the 3D case, $\sigma_l$ is more important.
Finally, in both figures, for a given scenario, the wider the beam is, the shorter the tail of the PDF becomes, and the smaller the maximum value of $h_g$, denoted by $A_0$ (which is denoted by dashed lines in Fig.~\ref{Fig:PDF}). This is in-line with our observations from Fig.~\ref{Fig:CondGML}.

\begin{figure*}[!tbp]
\begin{minipage}[t]{0.45\textwidth}
	\centering
		\includegraphics[valign=c,width=0.9\textwidth]{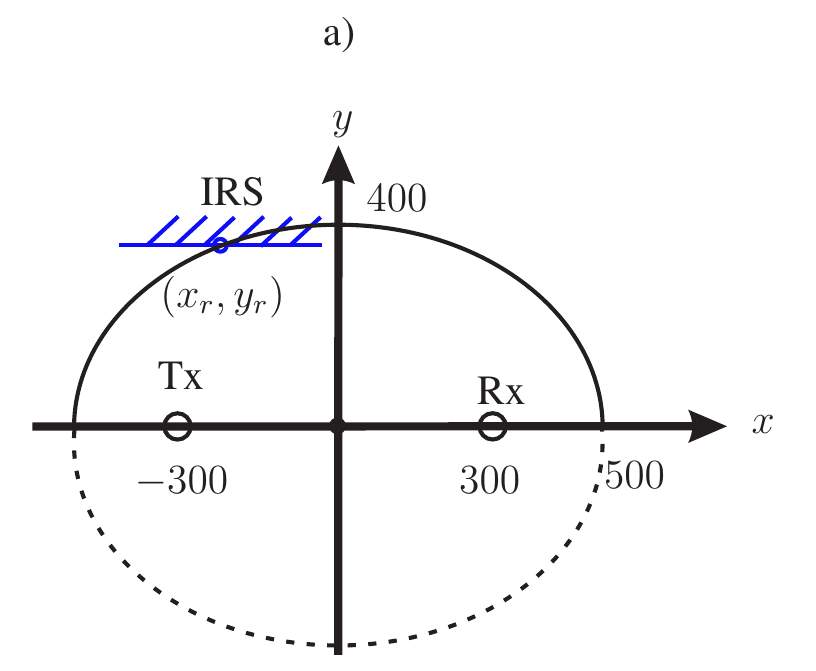}
	\end{minipage}
	\hfill
	\begin{minipage}[b]{0.1\textwidth}
	\end{minipage}
   \hfill
	\begin{minipage}[t]{0.5\textwidth}
		\centering
		\includegraphics[valign=c,width=1\textwidth]{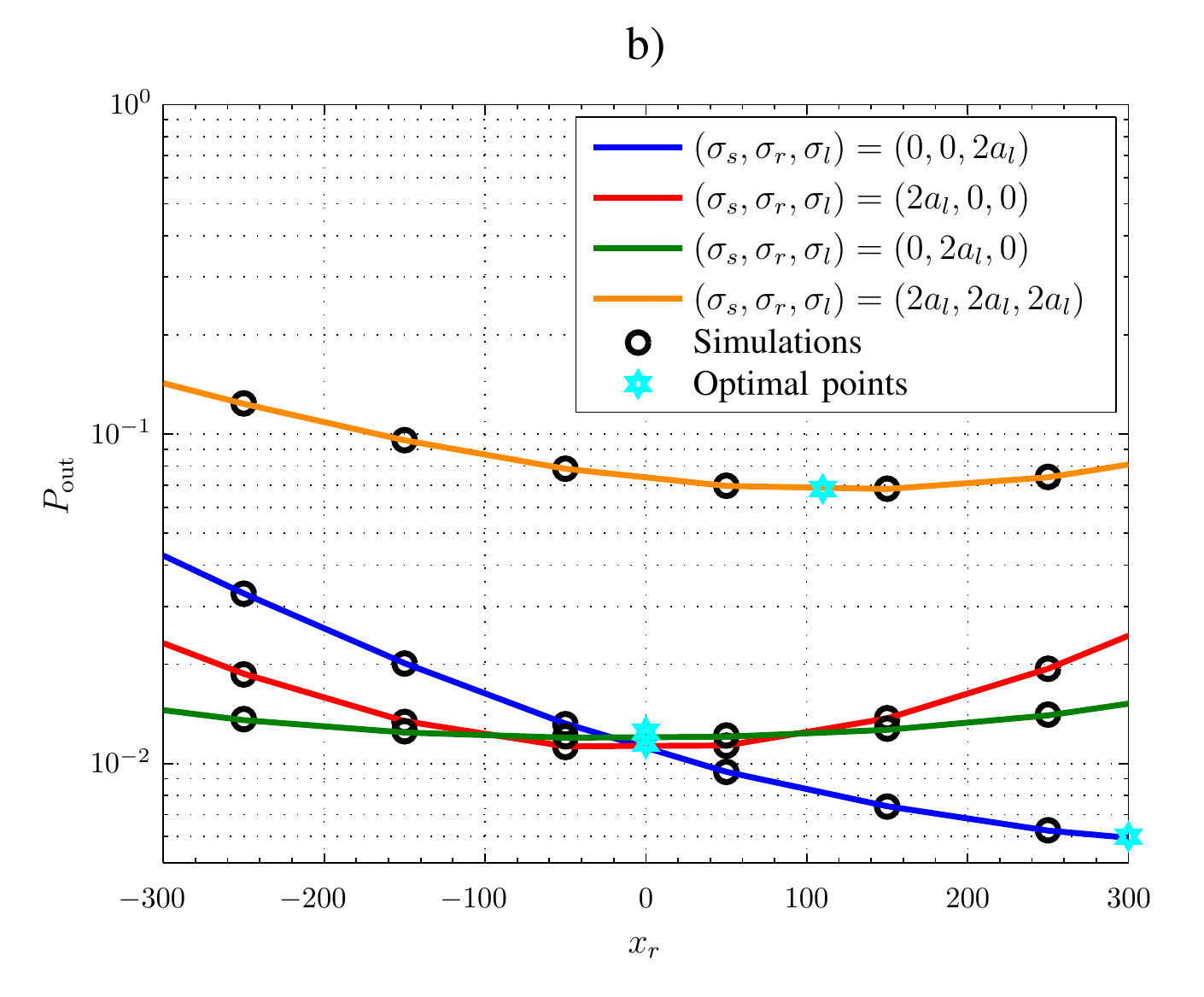}
	\end{minipage}
	\begin{minipage}[b]{0.1\textwidth}
	\end{minipage} \vspace{-0.3cm}
	\caption{a) Simulation setup for Fig.~\ref{Fig:Outage_Distance}b) and b) Outage probability  vs. $x_r$ for GG atmospheric turbulence ($C_n^2=1.7\times10^{-14}$~m$^{\frac{2}{3}}$), $d_{e2e}=1000$~m, and $\frac{w(d_{e2e},w_0)}{2a_l}=4$.}\vspace{-0.5cm}
	\label{Fig:Outage_Distance}
\end{figure*}

In Figs.~\ref{Fig:Outage_SNR} and \ref{Fig:Outage_Distance}, we plot the outage probability of an IRS-assisted FSO link for the 3D system. In particular, Fig.~\ref{Fig:Outage_SNR} shows $P_{\mathrm{out}}$ vs. the average received SNR  for two normalized beamwidths, namely $\frac{w(d_{e2e},w_0)}{2a_l}\in\{2,4\}$, and for LN and GG turbulence. From Fig.~\ref{Fig:Outage_SNR}, we observe that the simulation results match the analytical results for both LN and GG atmospheric turbulence. Moreover, the wider beam, i.e., $\frac{w(d_{e2e},w_0)}{2a_l}=4$, leads to a smaller $P_{\mathrm{out}}$ for high SNRs, since a wider beam is more robust against severe fluctuations. 

Finally, we study the impact of the position of the IRS. The simulation setup for Fig.~\ref{Fig:Outage_Distance}b) is depicted in Fig.~\ref{Fig:Outage_Distance}a). In particular, we assume that the Tx, IRS, and Rx are located in the $x-y$ plane at the same altitude. The location of the IRS changes on the upper perimeter of an ellipse with width $1000$~m and height $800$~m such that the end-to-end distance that the beam travels between the Tx and Rx, $d_{e2e}$, is constant, i.e., $d_{e2e}=1000$~m. Tx and Rx are located at the foci of the ellipse as shown in Fig.~\ref{Fig:Outage_Distance}a). Moreover, we assume that the IRS is always parallel to the $x-z$ plane, the lens plane is always orthogonal to the beam line, i.e., $\theta_{rl}=0$, and $\phi_r=\pi$ holds. This is in fact the special case studied in Corollary~\ref{corol:SpecialCase}.  Fig.~\ref{Fig:Outage_Distance}b) shows $P_{\mathrm{out}}$ vs. the location of the IRS, $x_r$, for an average received SNR of $20$~dB and for four fluctuation scenarios, namely Scenario~1: $(\sigma_s,\sigma_r,\sigma_l)=(0,0,2a_l)$, Scenario~2: $(\sigma_s,\sigma_r,\sigma_l)=(2a_l,0,0)$, Scenario~3: $(\sigma_s,\sigma_r,\sigma_l)=(0,2a_l,0)$, and Scenario~4:  $(\sigma_s,\sigma_r,\sigma_l)=(2a_l,2a_l,2a_l)$. We note that by increasing $x_r$, $\theta_i$ and $\theta_r$ become larger and smaller, respectively. Therefore, factors  $\frac{\cos\theta_r}{\cos\theta_i}$ and $\frac{\sin(\theta_i+\theta_r)}{\cos\theta_i}$ in Corollary~\ref{corol:SpecialCase} increase, which affects the following two system parameters: 
\textit{i)} The variances of the misalignment caused by the Tx  and the IRS building sway increase, cf. \eqref{Eq:Misalignment3D_specialcase}, which has a negative impact on $P_{\mathrm{out}}$, cf. Fig.~\ref{Fig:PDF}. Note that since it is assumed that the reflected beam is always orthogonal to the lens at the Rx, the variance of the misalignment caused by the Rx does not change when the IRS is moved. 
\textit{ii)} The width of the reflected beam becomes wider, cf. $w(d_{sr},\hat{w}_{0_1})$ in  Corollary~\ref{corol:SpecialCase}, which has a positive effect on $P_{\mathrm{out}}$ for high SNRs, cf. Fig.~\ref{Fig:Outage_SNR}. Therefore, for Scenario~1 where only the second effect exists, $P_{\mathrm{out}}$ decreases, as the IRS moves closer to the Rx. However, for Scenarios~2 and 3, due to the competing interplay of both effects, $P_{\mathrm{out}}$ first decreases and then increases as $x_r$ increases. Finally, for Scenario~4, the superposition of the misalignment caused by the Tx, IRS, and Rx leads to an optimal position for the IRS at $x_r=110$~m. By comparing Scenarios~1-3, we also observe that although the building sways
of the Tx, IRS, and Rx have identical variances, their relative impacts on $P_{\mathrm{out}}$ for a given position of the IRS  are not necessarily the same and depend on the relative positioning of the buildings, cf. \eqref{Eq:Misalignment3D_specialcase} in Corollary~\ref{corol:SpecialCase}. For example, when the IRS is closer to the Tx ($x_r\leq -100$~m), the building sway of the Rx has the largest impact and the building sway of the IRS has the smallest impact on  $P_{\mathrm{out}}$. 

\section{Conclusions and Future Research Directions}

In this paper, we proposed IRS-assisted FSO systems which can overcome the requirement of having a LOS between Tx and Rx. We analyzed 2D and 3D systems by studying equivalent mirror-assisted systems whose reflected electric fields are identical to those of the corresponding IRS-assisted systems employing the phase-shift profile proposed in this paper.  Based on these analyses, we developed conditional and statistical GML models which characterize the impact of the physical parameters of the IRS, such as its size, position, and orientation, on the end-to-end FSO channel. Moreover, we analyzed the outage probability of an IRS-assisted FSO link assuming that, in addition to the GML, atmospheric turbulence induced fading is also present. Simulation results validated the presented analysis and revealed important insights for system design. For instance, for a reasonable IRS size, e.g. $10$~cm for an end-to-end distance in the order of $1$~km, the beam truncation caused by the finite-size IRS can be safely neglected. 
Moreover, we showed via both simulations and theoretical analysis that even if the variances of the fluctuations of the Tx, IRS, and Rx positions are identical, their impact on the end-to-end channel is not necessarily the same and depends on the relative positioning of the three nodes. 

\Copy{Future}{ The analysis in this paper relies on tools from geometric optics and image theory and is valid only for the proposed IRS phase-shift profile. Thus, developing more general models based on the integral equation in \eqref{Eq:Huygens} and discrete phased-array IRS models, which are valid for general IRS phase-shift profiles, constitutes an important direction for future research. A recent preliminary study based on the integral equation in \eqref{Eq:Huygens} can be found in \cite{ajam2020channel}. Moreover, analysing the impact of phase-shift errors on the end-to-end channel gains for the phased-array IRS model is another interesting topic for future work, see \cite{Phase_Error_Badiu,Phase_Error_Vega} for similar analyses for IRS-assisted RF systems.}   In addition, the maximization of the system performance by optimization of the IRS phase-shift profile also merits further investigation.
Furthermore, we assumed that the IRS is able to only change the phase of the reflected beam, which was sufficient to realize anomalous reflection. However, sophisticated metasurfaces are able to modulate not only the phase but also the amplitude  and polarization of the incident wave \cite{OpticMetaSurf3,OMS_Tun}. Exploiting such capabilities to further improve the quality of IRS-assisted FSO links is yet another interesting topic for future work.

\appendices

\section{}\label{App:I_irs}

We use the fact that each point on the IRS is also located on another line, i.e., $\tilde{y}$, which is orthogonal to the beam line. Moreover, let $I^{\mathrm{irs}}_{\mathrm{G}}(y|d_{sr},\theta_i,w_0)\mathrm{d}y$ denote the fraction of power collected on the infinitesimally small line element $\mathrm{d}y$ around point $y$ on the IRS. Then, due to conservation~of~energy,~we~have
\begin{IEEEeqnarray}{lll}\label{Eq:2DI}
I^{\mathrm{irs}}_{\mathrm{G}}(y|d_{sr},\theta_i,w_0)\mathrm{d}y= I^{\mathrm{orth}}_{\mathrm{G}}\left(\tilde{y}|d,w_0\right)\mathrm{d}\tilde{y}, \quad\text{with}\quad \mathrm{d}\tilde{y}=\cos\theta_i\mathrm{d}y,
\end{IEEEeqnarray}
where $I^{\mathrm{orth}}_{\mathrm{G}}\left(\tilde{y}|d,w_0\right)\mathrm{d}\tilde{y}$ denotes the fraction of power collected on the infinitesimally small line segment $\mathrm{d}\tilde{y}$ around point $(\tilde{y},0)$ on a plane orthogonal to the beam line at distance $d$ from the LS. Moreover, from geometry, $d=d_{sr}+y\sin\theta_i$ and $\tilde{y}=y\cos\theta_i$ hold. However, since $d_{sr}\gg y\sin\theta_i$ holds in practice, we approximate $d\approx d_{sr}$. Therefore, \eqref{Eq:2DI} simplifies to 
\begin{IEEEeqnarray}{lll} \label{Eq:I_irs_I_orth_mid}
I^{\mathrm{irs}}_{\mathrm{G}}(y|d_{sr},\theta_i,w_0)&=\cos\theta_iI^{\mathrm{orth}}_{\mathrm{G}}\left(\cos\theta_i y|d_{sr},w_0\right)\nonumber\\
&=\frac{\sqrt{2}\cos\theta_i}{\sqrt{\pi} w(d_{sr},w_0)}\exp\Big(-\frac{2\cos^2\theta_i y^2}{w^2(d_{sr},w_0)}\Big)\overset{(a)}{=}I^{\mathrm{orth}}_{\mathrm{G}}\left(y|d_{sr},\tilde{w}_0\right),\quad
\end{IEEEeqnarray}
where  equation $(a)$ follows from comparing $\frac{\sqrt{2}\cos\theta_i}{\sqrt{\pi} w(d_{sr},w_0)}\exp\big(\frac{-2\cos^2\theta_i y^2}{w^2(d_{sr},w_0)}\big)$ with \eqref{Eq:PowerOrthogonal} and $\tilde{w}_0$ given in \eqref{Eq:w0new}. 
This completes the proof.

\section{}\label{App:Prop_TotalFrac}
To calculate the fraction of power that flows into the lens (i.e., the conditional GML), we first find the reflected power density across the lens and then integrate it over the lens. Similar to \eqref{Eq:I_irs_I_orth_mid} in Appendix~\ref{App:I_irs}, the reflected power density across the lens is obtained from the reflected power density across the plane perpendicular to the reflected beam, $I^{\mathrm{orth}}_{\mathrm{rfl}}(\cdot|\cdot,\cdot)$ given in \eqref{Eq:Truncated Gauss}, as $\cos\theta_{rl}I^{\mathrm{orth}}_{\mathrm{rfl}}(\rho\cos\theta_{rl}|d_{e2e},\hat{w}_0)$.
By integrating $\cos\theta_{rl}I^{\mathrm{orth}}_{\mathrm{rfl}}(\rho\cos\theta_{rl}|d_{e2e},\hat{w}_0)$ over the lens, the conditional GML~is~obtained~as
\begin{IEEEeqnarray}{lll} \label{Eq:Pt_App}
h_g=\cos\theta_{rl}\begin{cases}\displaystyle 
\int_0^{\rho_1}I^{\mathrm{orth}}_{\mathrm{rfl}}(\rho\cos\theta_{rl}|d_{e2e},\hat{w}_0)\mathrm{d}\rho+\int_0^{\rho_2}I^{\mathrm{orth}}_{\mathrm{rfl}}(\rho\cos\theta_{rl}|d_{e2e},\hat{w}_0)\mathrm{d}\rho,& \mathrm{if}\,\,\rho_{12}=2a_l\\
\displaystyle 
\Big|\int_0^{\rho_1}I^{\mathrm{orth}}_{\mathrm{rfl}}(\rho\cos\theta_{rl}|d_{e2e},\hat{w}_0)\mathrm{d}\rho-\int_0^{\rho_2}I^{\mathrm{orth}}_{\mathrm{rfl}}(\rho\cos\theta_{rl}|d_{e2e},\hat{w}_0)\mathrm{d}\rho\Big|,& \mathrm{otherwise,}
\end{cases}\quad
\end{IEEEeqnarray}
where $\rho_1$, $\rho_2$, and $\rho_{12}$ are given in Proposition~\ref{Prop:TotalFrac}. The two cases in \eqref{Eq:Pt_App} correspond to whether or not the center of the beam footprint lies on the lens line.
Substituting $\int_0^{\rho_i}\cos\theta_{rl}I^{\mathrm{orth}}_{\mathrm{rfl}}(\rho\cos\theta_{rl}|d_{e2e},\hat{w}_0)\mathrm{d}\rho
=\frac{1}{2}\mathrm{erf}\Big(\frac{\sqrt{2}\cos\theta_{rl}\rho_i}{w(d_{e2e},\hat{w}_0)}\Big)$ into \eqref{Eq:Pt_App} leads to \eqref{Eq:Pt} and concludes the proof.

\section{}\label{App:powdens_AG}

We first present the following lemma which is useful for derivation of the power density distribution across the IRS and later for derivation of the power density distribution across the lens in Appendix~\ref{App:GML_3D}.

\begin{lem}\label{Lem:tranfrofmPowDen}
Consider a $\tilde{x}\tilde{y}\tilde{z}$-coordinate system. Assume an LS is located at distance $\tilde{d}$ from the origin and its beam line strikes the origin with incident angle $\hat{\mathbf{\Psi}}_i=(\hat{\theta}_i,\hat{\phi}_i)$. Let $I^{\mathrm{orth}}(\hat{x}_0,\hat{y}_0|d)$ denote the power density distribution of the general beam (e.g., normal Gaussian and astigmatic Gaussian) at point $\hat{\mathbf{a}}=(\hat{x}_0,\hat{y}_0)^{\mathsf{T}}$ across a plane perpendicular to the beam line at distance $d$ from the LS. 
The power density distribution at point $\tilde{\mathbf{a}}=(\tilde{x}_0,\tilde{y}_0)^{\mathsf{T}}$ in the $\tilde{x}-\tilde{y}$ plane is given by
	\begin{IEEEeqnarray}{ccc}\label{Eq:Lem_tranfrofmPowDen}	
		I(\tilde{x}_0,\tilde{y}_0|\tilde{d},\hat{\mathbf{\Psi}}_i)=\cos \hat{\theta}_i I^{\mathrm{orth}}(\hat{x}_0,\hat{y}_0|\tilde{d})
				\quad \text{with} \quad
		\hat{\mathbf{a}}=\mathbf{T}_{\hat{\theta}_i}\mathbf{R}_{-\hat{\phi}_i}\tilde{\mathbf{a}}.\quad
	\end{IEEEeqnarray}
\end{lem}

\begin{IEEEproof}
We refrain from providing the complete proof due to space constraints and present only the basic idea of the proof in the following. Interested readers are referred to \cite[Appendix~D]{arXivIRS2020} for the complete proof.
 The term $\cos \hat{\theta}_i$ is essentially due to the conversation of energy such that the power flux across the  plane perpendicular to the beam line is identical to that across the non-perpendicular plane. Moreover, the transformation $\hat{\mathbf{a}}=\mathbf{T}_{\hat{\theta}_i}\mathbf{R}_{-\hat{\phi}_i}\tilde{\mathbf{a}}$ is used to determine the coordinates of point $\tilde{\mathbf{a}}$ in the $\hat{x}\hat{y}\hat{z}$-coordinate system on the perpendicular plane.
\end{IEEEproof}

Substituting $\hat{\mathbf{\Psi}}_i=(\theta_i,0)$ and \eqref{Eq:PowerOrthogonal} into Lemma \ref{Lem:tranfrofmPowDen} leads to \eqref{Eq:Iirslem3D} and completes the proof of Lemma~\ref{lem:PowerDistribution3D}.

\section{}\label{App:equivalence3D}
Using Lemma~\ref{Lem:tranfrofmPowDen} and \eqref{Eq:Iastigmatic}, the left-hand side of \eqref{Eq:I_irs_AG_I_irs_G} is obtained as
	\begin{IEEEeqnarray}{lll}\label{Eq:IastigApp} 
		I^{\mathrm{irs}}_{\mathrm{AG}}(x,y|d_{sr},\hat{\boldsymbol{\Psi}}_i,\hat{\mathbf{w}}_{0},\hat{\varphi}) 
		=  I^{\max}_{\mathrm{AG}}(d_{sr},\mathbf{w}_0) \cos\hat{\theta}_i \exp\big(\!\!-2\mathbf{a}^{\mathsf{T}}\mathbf{R}_{-\hat{\phi}_i}^{\mathsf{T}}\mathbf{T}_{\hat{\theta}_i}^{\mathsf{T}}\mathbf{R}_{\hat{\varphi}}^{\mathsf{T}}\mathbf{S}_{\hat{\mathbf{w}}_0}^{d_{sr}}\mathbf{R}_{\hat{\varphi}}\mathbf{T}_{\hat{\theta}_i}\mathbf{R}_{-\hat{\phi}_i}\mathbf{a}\big),
	\end{IEEEeqnarray}
where $\mathbf{a}=(x,y)^{\mathsf{T}}$ and $\mathbf{S}_{\hat{\mathbf{w}}_0}^{d_{sr}}=\mathrm{diag}\big\{\frac{1}{w^2(d_{sr},\hat{w}_{0_1})} ,\frac{1}{w^2(d_{sr},\hat{w}_{0_2})}\big\}$. Equating \eqref{Eq:IastigApp} and  \eqref{Eq:Iirslem3D}, we obtain
	\begin{IEEEeqnarray}{lll}\label{Eq:Matrixeq}
	\mathbf{R}_{-\hat{\phi}_i}^{\mathsf{T}}\mathbf{T}_{\hat{\theta}_i}^{\mathsf{T}}\mathbf{R}_{\hat{\varphi}}^{\mathsf{T}}\mathbf{S}_{\hat{\mathbf{w}}_0}^{d_{sr}}\mathbf{R}_{\hat{\varphi}}\mathbf{T}_{\hat{\theta}_i}\mathbf{R}_{-\hat{\phi}_i} = \mathbf{S}_{w_0}^{d_{sr}}\mathbf{T}_{\theta_i}^2 
	\Longrightarrow
	\mathbf{R}_{\hat{\varphi}}^{\mathsf{T}}\mathbf{S}_{\hat{\mathbf{w}}_0}^{d_{sr}}\mathbf{R}_{\hat{\varphi}}=\mathbf{A}.
	\end{IEEEeqnarray}
Here, $\mathbf{A}=\mathbf{T}_{\hat{\theta}_i}^{\mathsf{-T}}\mathbf{R}_{-\hat{\phi}_i}^{\mathsf{-T}}\mathbf{S}_{w_0}^{d_{sr}}\mathbf{T}_{\theta_i}^2\mathbf{R}_{-\hat{\phi}_i}^{-1}\mathbf{T}_{\hat{\theta}_i}^{-1}
\overset{(a)}{=}\mathbf{T}_{\hat{\theta}_i}^{\mathsf{-T}}\mathbf{R}_{\hat{\phi}_i}^{\mathsf{T}}\mathbf{S}_{w_0}^{d_{sr}}\mathbf{T}_{\theta_i}^2\mathbf{R}_{\hat{\phi}_i}\mathbf{T}_{\hat{\theta}_i}^{-1}$, where equality $(a)$ follows from $\mathbf{R}_{\tau}^{-1}=\mathbf{R}_{-\tau}$. Now, we have to find $\hat{\mathbf{w}}_0$ and $\hat{\varphi}$ such that \eqref{Eq:Matrixeq} holds.  In particular, since the columns of $\mathbf{R}_{\hat{\varphi}}$ are orthonormal and $\mathbf{S}_{\hat{\mathbf{w}}_0}^{d_{sr}}$ is a diagonal matrix, it is in the form of an eigenvalue decomposition. Therefore, the diagonal elements of $\mathbf{S}_{\hat{\mathbf{w}}_0}^{d_{sr}}$ are the eigenvalues of $\mathbf{A}$ and the columns of $\mathbf{R}_{\hat{\varphi}}^{\mathsf{T}}$ are the corresponding eigenvectors. Finally, we note that for the power density distributions in \eqref{Eq:IastigApp} and \eqref{Eq:Iirslem3D}, when the arguments of the exponential functions  become identical, the factors multiplied with them become identical as well. This completes the proof.

\section{}\label{App:GML_3D}

By substituting $\hat{\mathbf{\Psi}}_i=(\theta_{rl},0)$ and $d=d_{e2e}$ in Lemma \ref{Lem:tranfrofmPowDen} and \eqref{Eq:Iastigmatic}, we derive the power density across the lens at point $\mathbf{a}=(x,y)^{\mathsf{T}}$ as follows
\begin{IEEEeqnarray}{lll}\label{Eq:I_AG_Lens}
	I^{\mathrm{lens}}_{\mathrm{AG}}(x,y|d_{e2e},\hat{\mathbf{\Psi}}_i,\hat{\mathbf{w}}_0,\hat{\varphi})
	&=\cos\theta_{rl}I^{\max}_{\mathrm{AG}}(d_{e2e},\hat{\mathbf{w}}_0)\exp(-2\mathbf{a}^{\mathsf{T}}\mathbf{B}\mathbf{a})\nonumber \\
	&\overset{(a)}{=}\cos\theta_{rl}I^{\max}_{\mathrm{AG}}(d_{e2e},\hat{\mathbf{w}}_0)\exp(-2\mathbf{d}^{\mathsf{T}}\mathbf{\Lambda}\mathbf{d}),
\end{IEEEeqnarray}
where $\mathbf{B}=\mathbf{T}_{\theta_{rl}}^{\mathsf{T}}\mathbf{R}_{\hat{\varphi}}^{\mathsf{T}}\mathbf{S}_{\hat{\mathbf{w}}_0}^{d_{e2e}}\mathbf{R}_{\hat{\varphi}}\mathbf{T}_{\theta_{rl}}$. Equality $(a)$ follows from the eigenvalue decomposition of $\mathbf{B}$, where we have  $\mathbf{d}=\mathbf{U}^{\mathsf{T}}\mathbf{a}$. Moreover,  $\mathbf{U}$ contains the eigenvectors of $\mathbf{B}$ as its columns and $\mathbf{\Lambda}$ is a diagonal matrix with the corresponding eigenvalues on its main diagonal. Integration of \eqref{Eq:I_AG_Lens} over the lens area yields the power collected by the PD, which can be calculated using the same technique as in \cite[Eq. (13) and Appendix~C]{TCOM_2020}. This results in \eqref{Eq:hg_3D} and completes~the~proof.

\bibliographystyle{IEEEtran}
\bibliography{My_Citation_18-04-2021}

\end{document}